\documentclass[amssymb, amsmath, aps, pra, twocolumn]{revtex4-2}

\usepackage{hyperref}
\usepackage{xcolor}
\usepackage{bbm}
\usepackage{bm}
\usepackage{url}
\usepackage{graphicx}
\usepackage{amsmath,amsfonts,amssymb,nicefrac}

\usepackage{color}
\usepackage{upgreek}

\usepackage{xr}
\renewcommand{\vec}[1]{{\boldsymbol {#1}}}

\definecolor{rred}{HTML}{c04040}
\definecolor{ggreen}{HTML}{54c040}
\definecolor{bblue}{HTML}{4164c0}

\usepackage{mathtools}

\begin{document}
\title{Kinetic Event-Chain Algorithm for Active Matter}

\author{Nico Schaffrath}
\author{Thevashangar Sathiyanesan}%
\author{Tobias A. Kampmann}
 \author{Jan Kierfeld}%
 \email{jan.kierfeld@tu-dortmund.de}
\affiliation{%
Physics Department,  TU Dortmund University, 44221 Dortmund, Germany
}%


\begin{abstract}
   We present a cluster kinetic Monte-Carlo algorithm for active matter systems of 
    self-propelled particles with special focus on steric interactions. The kinetic 
    event-chain algorithm is based on the event-chain Monte-Carlo method and is applied
    to active Brownian disks in two dimensions. The algorithm assigns Monte-Carlo moves 
    of active disks a mean time based on a comparison between Brownian dynamics and the 
    dynamics of the event-chain Monte-Carlo method. This time is used to perform diffusional 
    rotation of their propulsion force. We show that the algorithm correctly and efficiently 
    reproduces various physical results ranging from single-particle dynamics to many-body-effects. 
    In particular, we reproduce  the phase diagram of active disks and
    the motility-induced phase separated region with high accuracy. The kinetic event-chain algorithm 
    is shown to be much faster - at comparable accuracy - than 
    (event-driven) Brownian dynamics algorithms, enabling large-scale simulations
     up to giant systems with $10^5$ particles  on standard  desktop hardware. 
\end{abstract}

\maketitle

\section{Introduction}

Active matter systems are non-equilibrium many-particle systems which are driven by 
a propulsion force at the level of individual particles. This gives rise  to continuous 
energy dissipation, local time-reversal symmetry breaking, and entropy 
production \cite{Bebon2024}. Active matter can be found at all length scales both 
for synthetic and biological systems \cite{Marchetti2013}; examples comprise self-propelled 
colloidal particles \cite{bechinger2016} and microswimmers \cite{Elgeti2015}, 
cytoskeletal filaments propelled by molecular motors \cite{Surrey2001,Kruse2004,Kraikivski2006}, 
colonies of bacteria \cite{wu2000}, and swarms of animals \cite{Vicsek2012}. 

In the simplest active systems, the propulsion force direction 
is anchored to the individual particle, which  breaks the rotational 
symmetry at the particle level. The force changes direction 
upon diffusive re-orientation of the particle. 
The propulsion energy sets particles in motion and is dissipated 
via a surrounding bath, which  breaks detailed balance and gives rise to
non-equilibrium physics. The  most popular theoretical and 
numerical model are active Brownian
particles (ABPs) with an overdamped motion and a heat bath 
modeled by Brownian random forces (Langevin dynamics).

Interacting active matter exhibits novel collective phenomena, 
which are not present in the corresponding equilibrium systems. 
One prominent example is the motility-induced phase 
separation (MIPS) of self-propelled hard spheres or disks \cite{Bialke2013,Cates2015},
which is an activity-driven clustering phenomenon driven by the active
blocking of particles with opposing propulsion directions. 
More complicated phase behavior occurs for non-spherical
particles or in the presence of alignment interactions \cite{Marchetti2013}. 

Despite considerable progress during the last decade \cite{Shaebani2020},
simulation methods for active many-particle systems are largely limited to
Langevin or molecular dynamics techniques. While Monte-Carlo (MC) algorithms produced
much of the numerical progress in statistical physics 
of equilibrium phase transitions, the development of 
kinetic MC schemes for 
active matter started only recently
\cite{levis2014,klamser2018,klamser2019,Klamser2021}. 
Moreover, cluster MC algorithms have not been proposed at all 
for active systems so far, while rejection-free cluster MC algorithms have 
proven to be a powerful tool for lattice spin 
systems \cite{Swendsen1987,Wolff1989}  
and -- as event-chain (EC) MC \cite{bernard2009} -- also
for off-lattice simulations of many-particle systems.

\begin{figure}
	\includegraphics[width=0.875\linewidth]{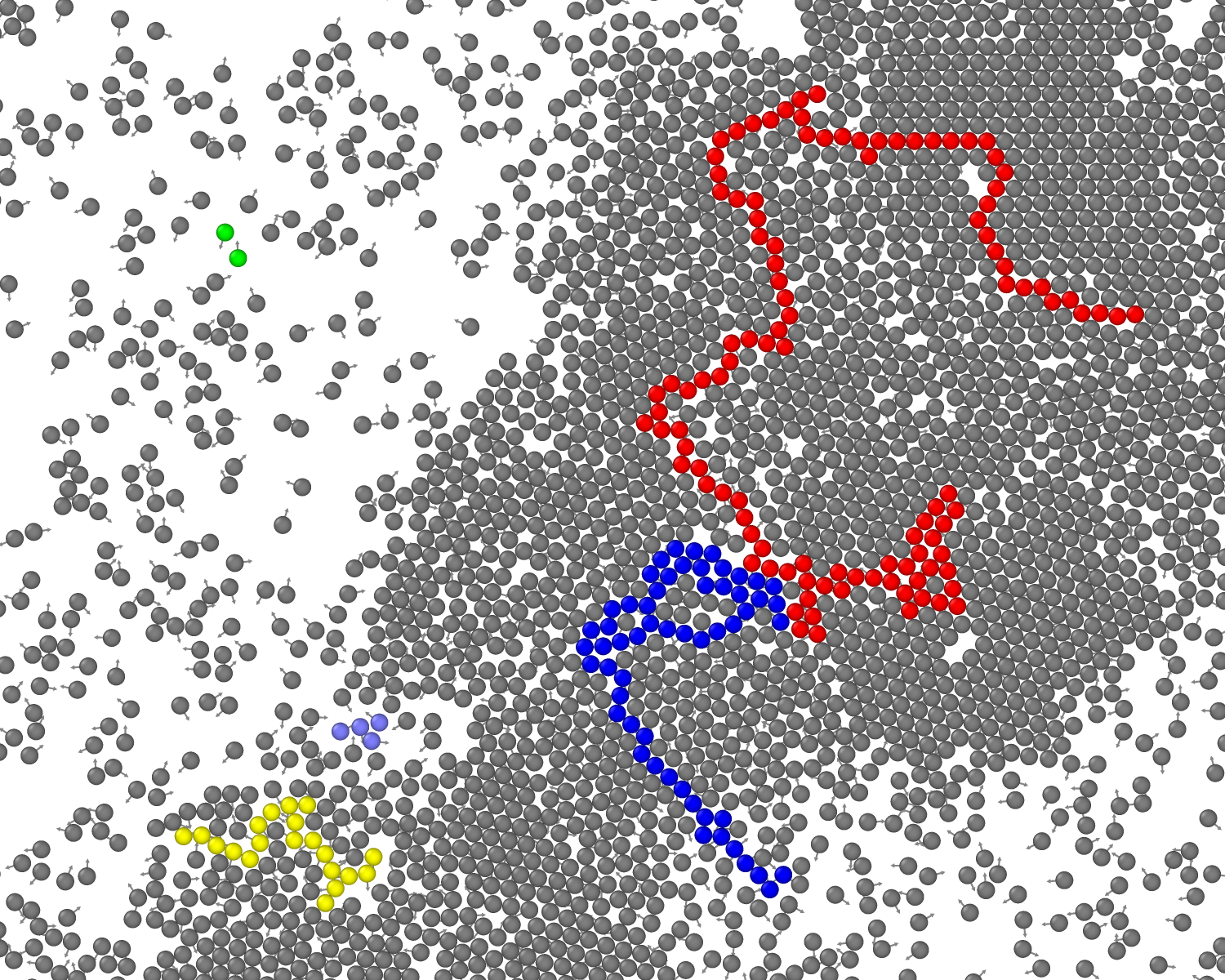}
	\caption{
        Exemplary snapshot of a system of hard ABPs where MIPS occurs. 
        To each particle a 
        small arrow is anchored, representing the current direction of the active force. The 
        particles, which participated in the last five cluster moves are highlighted using 
        colors. The higher the local packing fraction, the more particles are displaced in 
        a typical cluster move.
    }
	\label{fig:kEC_example_cluster_move_part_1}
\end{figure}

In this paper, we present the first cluster kinetic MC scheme for 
the simulation of active matter systems of self-propelled particles, with 
main focus on steric interactions, i.e., genuinely hard particles.
The algorithm is a kinetic event-chain (kEC) algorithm based on the ECMC 
method.
A dense system of ABPs with exemplary MC cluster moves is depicted in 
Fig.\ \ref{fig:kEC_example_cluster_move_part_1}. 
Throughout this work, we use OVITO \cite{ovito} to visualize configurations of interest. 
The algorithm becomes exact in the limit of small EC lengths, i.e., 
small MC cluster moves, and is applied to 
systems of genuinely  hard active disks and 
active Lennard-Jones disks in two dimensions.
It is important to note that standard Brownian dynamics simulations 
can not be applied to hard active particles. 
We verify the novel kEC algorithm by analyzing multiple observables
in these systems, particularly focusing on the 
phase diagram of active hard disks and mapping out the MIPS region.
For Lennard-Jones disks, we can compare to Brownian dynamics simulations performed using the 
LAMMPS package \cite{LAMMPS}, with respect to accuracy. 
For hard disks, we can benchmark and compare to a version of the 
Event-Driven Brownian-Dynamics (EDBD) algorithm \cite{scala2007event}, which 
takes the self-propulsion of ABPs into account \cite{Ni2013} and was used 
by Levis \textit{et al.} \cite{Levis2017} to investigate the systems MIPS region.

\section{Kinetic event-chain Monte-Carlo algorithm}

EC algorithms are a variant of irreversible Markov-Chain MC 
algorithms \cite{Krauth2021} and  have been applied to a wide
range of systems in the last decade 
\cite{bernard2009,Michel2014,michel2015,kampmann2015,nishikawa2016,harland2017,Klement2019,kampmann2021}. They can also be classified as cluster algorithms. 
We focus on steric interactions to introduce  the concepts 
of the algorithm in the following.
In a hard disk system, an EC cluster is constructed by choosing 
a random unit direction $\vec{e}_\text{EC}$ and a random initial disk.
Starting with the initial disk, the ``moving'' disk is displaced 
in direction $\vec{e}_\text{EC}$ until collision with another disk. 
In a billiard-like fashion, the hit disk becomes the ``moving'' 
disk (in a so-called \emph{lifting move}) and is displaced 
in direction $\vec{e}_\text{EC}$ until the next collision. 
The cluster move ends if the total displacement 
equals a prescribed EC length $\ell_\text{EC}$, which is a 
parameter of the EC algorithm. The resulting cluster moves 
of chains of particles are rejection-free and satisfy global 
balance rather than detailed 
balance \cite{bernard2009,michel2015}. Compared to local 
Metropolis-Hastings MC \cite{metropolis1953}, speed-ups of up to 
two orders of magnitude can be achieved 
\cite{bernard2009,kampmann2015,Klement2019}.

MC algorithms are not based on an equation of motion and, 
thus, lack a proper definition of time. ABPs, on the other hand, 
are described by explicitly time-dependent Brownian dynamics. 
A system of $N$ active particles (particle index $i$) in two dimensions, 
which are driven with propulsion force~$F_0$ in the unit directions 
$\vec{e}_i(t)=\left( \cos(\theta_i), \sin(\theta_i)\right)$,
is described by 
\begin{equation}
    \dot{\vec{r}}_i(t) = v_0 \vec{e}_i(t) + \vec{\xi_i}(t) + \vec{f_i}(t),
    \label{eq:ABP1}
\end{equation}
where~$\vec{r}_i(t)$ are the particle positions. 
Here, $v_0 = F_0/\Gamma$ is the driving 
velocity for a friction constant $\Gamma$, and $\vec{\xi_i}(t)$ 
a Gaussian thermal noise with zero mean and 
correlations~$\langle \vec{\xi_i}(t) \vec{\xi}_{j}^T(t')\rangle = 2D 
\vec{1} \delta_{ij}\delta(t-t')$, where $D= k_BT/\Gamma$ 
is the translational diffusion constant. 
The remaining active particles exert interaction
forces $\vec{f_i}= \vec{F}_i/\Gamma$; we focus on 
monodisperse systems with particle 
diameter $\sigma$. The propulsion direction $\vec{e}_i(t)$ 
undergoes free rotational diffusion, i.e., 
\begin{equation}
    \langle \dot{\theta}_i(t) \dot{\theta}_j(t') \rangle = 2 D_r \delta_{ij} \delta(t-t'),
     \label{eq:ABP2}
\end{equation}
with the rotational diffusion constant $D_r = 3D /\sigma^2$ (for hydrodynamic reasons 
\cite{Levis2017}). The activity of a single ABP is characterized by the dimensionless 
P\'{e}clet number 
\begin{equation}
    \mathrm{Pe} \equiv \frac{v_0}{\sigma D_r}.
\end{equation} 

Individual active particles perform persistent random walks 
with a persistence time $\tau  =  1/D_r$ 
because the propulsion force direction undergoes rotational diffusion. Equilibrium MC algorithms
lack this persistent motion because of their Markovian nature. 
Several kinetic MC schemes have been proposed for active particles
\cite{levis2014,klamser2018,klamser2019,Klamser2021}, where a temporal correlation 
of MC particle moves has been introduced to establish persistence. 
In conjunction 
with Metropolis sampling of interactions, these kinetic MC schemes 
show non-equilibrium phenomena like MIPS for repulsive active particles. 
For example, the algorithm developed by Klamser \textit{et al.} \cite{Klamser2021} advances 
time in equidistant timesteps $\Delta t$ and the direction $\vec{e}_i$ is updated 
by a small random rotation ensuring the proper persistence. Regarding the spatial
displacement the authors showed, both numerically and analytically, that 
their method becomes ill-defined in the limit of small timesteps $\Delta t$ if active particle 
displacements rely purely on persistent steps. To restore correct dynamics, a blended
algorithm was introduced, which mixes steps in random directions to the persistent ones.
For the blended algorithm  an exact mapping to ABPs can be established, as the authors subsequently showed
their method leads to the same Fokker-Planck equation in the limit of small timesteps 
$\Delta t$ as Brownian dynamics.

The very basic idea of our kEC algorithm, in order to simulate ABPs, is to use the standard
ECMC method to generate moves, based on the rules for passive equilibrium systems, and
afterwards update the direction of the active force using a newly introduced mean time per move.
This way, we want to achieve an algorithm that is as effective as ECMC simulations for hard spheres, while it can 
correctly deal with the dynamic situation of active particles.

With this in mind, we view each active particle as a particle under its individual constant
force $F_0 \vec{e}$ and thus, in its individual constant linear potential over the course of
each ECMC move. 
We then calculate the mean move length in force direction $\langle \Delta \vec{r}_\text{EC} 
\cdot \vec{e} \rangle$ of a single particle per ECMC move (see below for the analytical 
calculation). Because $\langle \Delta \vec{r} \cdot \vec{e} \rangle= v_0 \Delta t$ holds for 
a single ABP exactly, we identify $\langle \Delta t_\text{EC} \rangle = \langle \Delta
\vec{r}_\text{EC} \cdot \vec{e} \rangle / v_0$.
By assigning $\langle \Delta t_\text{EC} \rangle$ to pass for each ECMC move, we are able
to introduce an exact mean time into our algorithm. 
It amounts to a dynamical mean-field assumption, because we apply a mapping 
between kEC time and ABP time that has been derived from the mean time for 
an ECMC displacement move.
Because we also find $\langle \Delta \vec{r}_\text{EC} \cdot \vec{e} \rangle 
= \langle \Delta \vec{r}_{\mathrm{EC}} \rangle\cdot \vec{e}$ for the ECMC algorithm,
we obtain the identical time assignment rule based on the mean displacement 
vector $\langle \Delta \vec{r}_{\mathrm{EC}} \rangle$
per ECMC move. This will be important later on to derive that 
the kEC algorithm becomes exact in the limit of small EC lengths.
Using the assigned time $ \langle \Delta t_\text{EC}\rangle$, 
we re\-adjust the force direction according 
to rotational diffusion after each ECMC move.

For the analytical calculation of the mean move 
length in force direction 
$\langle \Delta \vec{r}_\text{EC} \cdot \vec{e} \rangle$, 
it is sufficient to consider a single active particle 
under a constant force $F_0 \vec{e}$ or in a 
corresponding linear potential $V(\vec{r})$.
In an external potential, the ECMC algorithm works as follows
\cite{Michel2014,harland2017,kampmann2021}: 
A random EC direction $\vec{e}_\text{EC}$ is 
chosen. If the move is energetically downhill 
($\cos\phi \equiv \vec{e}_\text{EC}\cdot \vec{e}\ge 0$), 
the particle is displaced by the full EC length, 
$\Delta \vec{r} = \ell_\text{EC}\vec{e}_\text{EC}$. 
If the move is energetically uphill ($\cos\phi < 0$), a ``usable'' energy 
$\Delta U>0$ is drawn from a Boltzmann distribution 
$p(\Delta U) \sim \exp(-\beta \Delta U)$ (with $\beta = 1/k_B T$), 
and a rejection distance $d_\text{rej}$ is determined
from $\Delta U = -F_0 d_\text{rej}\cos\phi$. 
If $d_\text{rej}\ge\ell_\text{EC}$, the particle is moved 
by the full EC length, $\Delta \vec{r} = \ell_\text{EC}\vec{e}_\text{EC}$. 
If $d_\text{rej}<\ell_\text{EC}$, the
particle is only moved by 
$\Delta \vec{r} = d_\text{rej}\vec{e}_\text{EC}$, the EC direction 
is lifted $\vec{e}_\text{EC} \to \vec{e}_\text{EC}'$ 
(by reflecting with respect to the equipotential surface, i.e.,
$\vec{e}_\text{EC}' = \vec{e}_\text{EC} - 2 \cos\phi \vec{e}$),
and the particle is moved by the remaining EC length
$\ell_\text{EC}-d_\text{rej}$ in the new
downhill direction. In total, this results in 
$\Delta \vec{r} =d_\text{rej} \vec{e}_\text{EC}+
(\ell_\text{EC}-d_\text{rej}) \vec{e}_\text{EC}'$. 
After this EC move, a new direction 
$\vec{e}_\text{EC}$ is chosen, and we start over. 
Averaging over all cases and all randomly
chosen EC directions $\vec{e}_\text{EC}$ (angles $\phi$), 
we obtain 
\begin{equation*}
   \frac{\langle \Delta \vec{r}_\text{EC} \cdot \vec{e} \rangle}{\ell_\text{EC}}  = x f(x)
\end{equation*}
with a dimensionless parameter 
 \begin{equation}
       x\equiv \beta F_0 \ell_\text{EC} = v_0 \ell_\text{EC} / D =
         3\mathrm{Pe} \ell_{\mathrm{EC}} / \sigma 
       \label{eq:x}
 \end{equation}
and a 
scaling function
\begin{equation}
	f(x) = \frac{2}{\pi} \frac{1}{x} -
	\frac{1}{x^2} \frac{1}{\pi} \int_{-\pi/2}^{\pi/2} d\phi
	(1- e^{-x\cos\phi}),
	\label{eq:fx}
\end{equation}
which is monotonically decreasing 
with asymptotics $f(x) \approx 1/4$ for $x\ll 1$ and $f(x) \approx 2/\pi x$ for $x\gg 1$, 
see Supplemental Materials (SM) \cite{SM}, Sec.\ \ref{sec:meantEC}).

For the ECMC move mean time $ \langle \Delta t_\text{EC}
\rangle = \langle \Delta \vec{r}_\text{EC} \cdot \vec{e} \rangle / v_0$, 
we use $\ell_\text{EC}/v_0 = \tau x/3\mathrm{Pe}^2$ and obtain 
\begin{equation}
    \langle \Delta t_\text{EC} \rangle
    = \tau \frac{1}{3\mathrm{Pe}^2} x^2f(x) 
    \label{eq:tEC}
\end{equation}
such that $\langle \Delta t_\text{EC}\rangle$ only depends on the active force $F_0$ and 
the EC length $\ell_\text{EC}$ via $\mathrm{Pe}$ and $x$. 
For weak active drive or small EC lengths $x\ll 1$, we get  $\langle
\Delta t_\text{EC} \rangle \approx   \ell_\text{EC}^2 / 4D$ reminiscent of a diffusive time scale. 
The opposite limit
of strong active drive or large EC lengths $x \gg 1$ leads to
$\langle \Delta t_\text{EC}\rangle \approx  2\ell_\text{EC}/\pi v_0$ reminiscent of a time scale for directed motion. 

In the simulation, after each ECMC move, the force direction 
is updated $\theta' = \theta + \Delta \theta$ by drawing an angle $\Delta \theta$ from a
Gaussian distribution using~$\mu_{\theta} = \theta$ and~$\sigma_{\theta} = \sqrt{2 D_r 
\langle \Delta t_{\mathrm{EC}} \rangle}$. For a single active particle, this is only a good
approximation if updates remain smooth, $\langle \Delta t_\text{EC} \rangle\ll \tau$. 
From  (\ref{eq:tEC}), we can show  this is equivalent to 
$x\ll (3\pi /2) \mathrm{Pe}^2$ or to require moderately small EC lengths 
$\ell_\text{EC}/\sigma \ll (\pi /2) \mathrm{Pe}$ for high activity $\mathrm{Pe}>1$ 
(and $x\ll 2\sqrt{3} \mathrm{Pe}$  or $\ell_\text{EC}/\sigma \ll 2/\sqrt{3}$ in the passive limit  $\mathrm{Pe}\ll 1$).  

In the many-particle system, we combine the ECMC procedure 
for one active particle with the 
hard disk collision ECMC rules outlined above. 
Then, we simulate  a Hamiltonian~${\cal H} =
\sum_i[- F_0 \vec{e}_i\cdot \vec{r}_i +V_i(\{\vec{r}_i\})]$, 
where the driving  forces~$F_0 \vec{e}_i$ enter as individual 
external linear potentials, with the usual ECMC 
scheme assuming fixed~$\vec{e}_i$ during each EC move.

This results, as a consequence of the previously mentioned 
\emph{lifting moves}, in ECs with 
more than one participating particle. 
The purpose of lifting  is to avoid MC rejections caused by 
collisions by lifting the event chain to the collision partner 
and continuing the MC motion. As a result, in each EC, 
we move particles 
by the same total distance $\ell_\text{EC}$ as in a single
particle system, but distribute the total distance to all particles 
$i$ in the EC, which move a distance $\Delta d_i$ with 
$\sum_{i\in EC}\Delta d_i  = \ell_\text{EC}$.
In order to assign a proper 
time to each moving particle $i$ in the EC, we distribute the 
mean time~$\langle \Delta t_\text{EC} \rangle$ of a single
particle EC move without collisions over all particles~$i$ 
in the EC according to their
displacements~$\Delta d_i$. Since collisions are instantaneous 
in case of hard disks this leads to 
\begin{equation}
    \langle \Delta t_\text{EC} \rangle_i = \frac{\Delta d_i}{\ell_\text{EC}} \langle \Delta t_\text{EC} \rangle.
    \label{eq:tECN}
\end{equation}
After each EC, the force directions of all participating particles~$i$ 
are updated by drawing $\Delta \theta_i$ from Gaussian distributions
with mean $\mu_{\theta_i}=\theta_i$  and variance 
$\sigma_{\theta_i}^2=2 D_r \langle \Delta t_{\mathrm{EC}} \rangle_i$.
The assignment of individual times $\langle \Delta t_\text{EC} \rangle_i$ leads to 
each particle having its ``own simulation time''. 

The choice (\ref{eq:tECN}) for the time $\langle \Delta t_\text{EC} \rangle_i$ proportional to its moving distance $\Delta d_i$ in the EC is the only choice which ensures a \emph{constant} algorithmic particle velocity $v_i = \Delta d_i / \langle \Delta t_\text{EC} \rangle_i$
(regarding collisions and 
subsequent lifting event to be instantaneous)
and, thus, a uniform  particle current along an EC. 
Within each  EC global balance is satisfied by construction of the algorithm. 
If ECs are started where the previous EC ended, this results in strict global balance.
This means that the $N$-particle probability current density  in configurational space is divergence-free. 
Starting ECs where the previous EC ended can produce artifacts in the distribution of the individual particle simulation times, 
because (i) particles in dense regions tend to be updated not only proportional to density, but more frequently and (ii) spatially inhomogeneities of this distribution are amplified. 

Therefore, we start new ECs at  randomly chosen particles but the algorithm should retain divergence-free probability current densities on average. 
With the choice (\ref{eq:tECN}) translating configurational moves into moves with constant velocity this leads to a divergence-free $N$-particle current density. 
This is, however, exactly the condition for a dynamic stationary state in a non-equilibrium system such as active disks \cite{Zia2007}.
Therefore, the choice (\ref{eq:tECN}) is the unique choice that is 
compatible with a dynamic stationary state in the active system.  In (\ref{eq:tECN}) we also use  the  single
particle EC move time $\langle \Delta t_\text{EC} \rangle$ in the absence of  collisions to be distributed over the over all particles~$i$. This guarantees correct currents in \emph{dilute} regions of a system, where particles are essentially isolated. Ensuring constant currents along ECs that can cross the entire system 
by using (\ref{eq:tECN})  
then correctly establishes the necessary current equilibrium between dense and dilute regions of the system.  
Loosely speaking, (\ref{eq:tECN})  enforces the kEC currents to be divergence-free and constant along an EC 
via global balance, and the size of the currents generated by the kEC algorithm is gauged to be correct in the dilute regions.

As stated above, the assignment of individual times $\langle \Delta t_\text{EC} \rangle_i$ according to Eqs.\ 
 (\ref{eq:tEC}) and  (\ref{eq:tECN}) leads to 
each particle having its ``own simulation time''.  Therefore, we have to make sure that each particle has essentially 
the same simulation time, i.e., that the distribution of 
particle simulations times is sufficiently narrow.
We confirmed that the relative standard 
deviation $\sigma_{\Delta t}$ of the resulting distribution of simulation times $\Delta t$  
decreases continuously by increasing the 
mean simulation time $\mu_{\Delta t}$ per particle or by decreasing 
the EC length, $\sigma_{\Delta t}/\mu_{\Delta t} = a(N)\ell_\text{EC}^\alpha \mu_{\Delta t}^{-1/2}$,
as detailed in the SM \cite{SM}. This holds for different 
values of parameters, i.e., regardless of whether MIPS occurs. Therefore,
the algorithm can be set up so that the simulation time is, 
on average, equal for all particles. 
A condition  $\sigma_{\Delta t} < \tau$ will  assure that particles remain within 
one persistence time, i.e., the propulsion directions will not be affected by 
the time spread. For $\ell_{\mathrm{EC}} =0.01\sigma$, we can 
simulate for several thousand
persistence times using the kEC algorithm
without  violating this criterion.
Also the above condition $\langle \Delta t_\text{EC} \rangle D_r \ll 1$ 
for smooth rotations becomes much weaker 
in a many-particle system, as it only requires
$\langle \Delta t_\text{EC} \rangle_i D_r \ll 1$ and 
$\langle \Delta t_\text{EC} \rangle_i\ll \langle \Delta t_\text{EC} \rangle$.
This is hardly ever a concern.

\section{Validity of the algorithm}

\begin{figure}
    \includegraphics[width=1\linewidth]{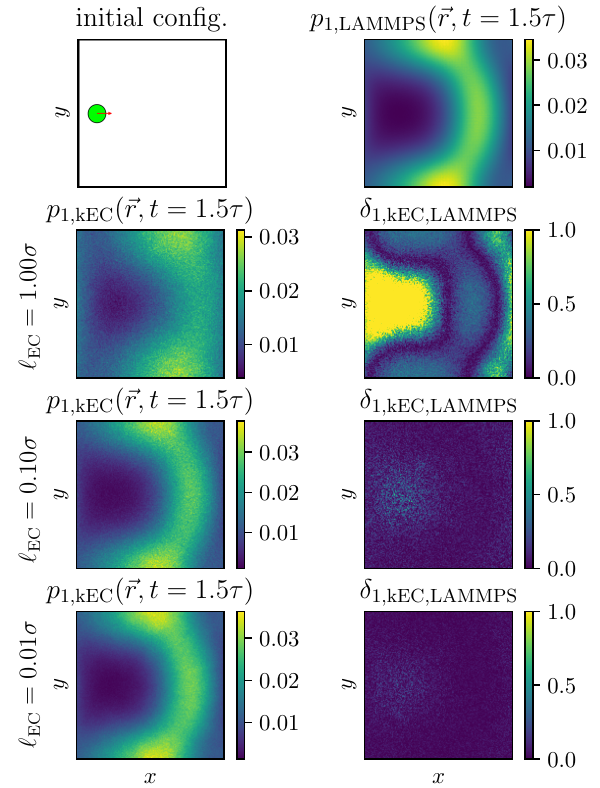}
    \caption{
        Probability density distribution $p_1(\vec{r},t)$ 
        of a single ABP ($\mathrm{Pe} = 4$) starting 
        from a given initial condition with periodic boundary 
        conditions (see upper left) at $t = 1.5\tau$ 
        using a LAMMPS Brownian dynamics simulation 
        and the kEC algorithm as well as the absolute value of the 
        relative deviation between both algorithms per grid element 
        $\delta_{\mathrm{1,kEC,LAMMPS}} \equiv \vert (p_{1,\mathrm{kEC}} -
        p_{1,\mathrm{LAMMPS}}) / p_{1,\mathrm{LAMMPS}} \vert$. 
        The results show an increase 
        in accuracy of the kEC algorithm for decreasing EC length 
        supporting  the claim that the kEC algorithm becomes exact 
        in the limit~$\ell_{\text{EC}} / \sigma \rightarrow 0$.
        }
    \label{fig:one_body_problem_probability_density_distribution}
\end{figure}

\begin{figure}
    \includegraphics[width=1\linewidth]{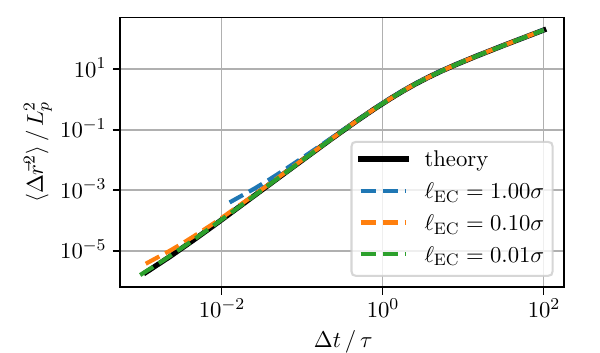}
    \caption{
    	Comparison between the kEC algorithm and the 
    	analytical solution for  the mean square displacement 
    	$\langle \Delta \vec{r}^2 \rangle$ of a single free ABP ($\mathrm{Pe} = 50$)
    	as a function of time $\Delta t/\tau$, 
    	for different EC lengths $\ell_{\text{EC}}$. Reducing 
    	the EC length improves the accuracy of the algorithm, 
    	especially on short timescales.
    	}
    \label{fig:one_body_problem_msd}
\end{figure}

Our above arguments suggest that the kEC algorithm samples single 
active particles (the dilute limit) 
correctly on average  
and that it samples stationary dynamic states of many active particles correctly on average. This 
does not mean that the microscopic dynamics will be exact. 
The kEC algorithm can become potentially inexact because of the effective time-dependence 
of the Hamiltonian: the external driving forces represent individual external linear 
potentials, but they change direction as a function of time by rotational diffusion. 
We handle this time-dependence by introducing the mean time $\langle \Delta t_\text{EC} \rangle$ per EC for a single ABP and redistributing this time properly over the particles participating in an EC for many ABPs. This procedure 
is only correct \emph{on average}. 
However, there are two opposite limiting cases, where our algorithm  becomes
strictly exact: (i) in the passive limit $\mathrm{Pe} \to 0$, 
the driving potentials effectively vanish and (ii) in the opposite limit $\mathrm{Pe} 
\to \infty$ of large P\'{e}clet numbers $\mathrm{Pe} \equiv v_0 / (\sigma D_r)$ or 
$D_r\to 0$ the directions of the driving forces become essentially quenched. 
In both cases the Hamiltonian becomes effectively time-independent, the correctness 
of the  mean time $\langle \Delta t_\text{EC} \rangle$ per EC irrelevant, and 
the kEC algorithm equivalent to a standard ECMC simulation.

The exactness both for 
$\mathrm{Pe} \to 0$ and $\mathrm{Pe} \to \infty$ already suggests that the kEC algorithm 
should stay accurate over a fairly large range of P\'{e}clet numbers and 
deviations for 
all intermediate P\'{e}clet numbers remain small. For intermediate 
 P\'{e}clet numbers, the dimensionless parameter $x=\beta F_0 \ell_\text{EC}$ from Eq.\ (\ref{eq:x})
plays a key role for the validity of our algorithm. This can be shown  explicitly for a single 
ABP.  We assign a kEC simulation time $\langle \Delta t_\text{EC} \rangle$ to a single ABP  
using  the \emph{first} moment of 
the displacement in force direction
$\langle \Delta \vec{r} \cdot \vec{e} \rangle= v_0 \Delta t$  by requiring that 
$\langle \Delta \vec{r}_\text{EC} \cdot \vec{e} \rangle= v_0 \langle \Delta t_\text{EC} \rangle$.
As shown in the SM \cite{SM}, 
Sec.\ \ref{sec:validation_sm}), we can also calculate \emph{higher} moments  
related to the dynamics of a single free ABP, such as $\langle (\Delta \vec{r}_\text{EC} 
\cdot \vec{e})^2 \rangle$ and $\langle \Delta \vec{r}_\text{EC}^2 \rangle$, on a single-step 
level. However, these quantities only match their analytical counterparts from Brownian dynamics 
in the limit $x \rightarrow 0$. On the contrary: the larger $x$ 
the more does our algorithm's single-step behavior differ from actual Brownian dynamics of ABPs.  
More importantly, we derive a Fokker-Planck equation for the probability density $P_{\mathrm{EC}}(\vec{r},
\theta, t)$ to find a single particle at position $\vec{r}$ and force direction $\theta$ at times $t$
for the  dynamics of the kEC algorithm in the  SM \cite{SM}, Sec.\ \ref{sec:validation_sm}).  
The Fokker-Planck equations contains a term for angular diffusion, a convective term for directed translational motion 
and a positional diffusion term, which can be compared with their counterparts in the Fokker-Planck equation 
for a single ABP.
We  show that (i) the angular diffusion term agrees in the limit $\langle \Delta t_\text{EC} \rangle \ll 1/D_r = \tau$;
(ii) the convective term agrees with exactly our choice  $\langle \Delta \vec{r}_{\mathrm{EC}}\rangle =
        \vec{v}_0  \langle \Delta t_\text{EC} \rangle$  of the time assignment; 
        (iii) the positional diffusion term agrees in the 
        aforementioned limit $x \rightarrow 0$.
This is an important result, because it establishes a limit where the kEC algorithm becomes 
exact on the single particle level. 
Further inspection of the positional diffusion term in the Fokker-Planck equation 
shows that we can increase $x$ such that 
$\ell_\text{EC}/\sigma$ becomes of  order unity 
if we are not interested in features of the distribution  $P(\vec{r}, \theta, t)$ 
below one particle diameter $\sigma$.

It may be worth noting the continuous time limit of our algorithm is not affected by the 
limitations found by Klamser \textit{et al.} \cite{Klamser2021} mentioned above, mainly because 
our algorithm does not rely on purely persistent steps by default, as we choose the 
displacement direction $\vec{e}_\text{EC}$ at the beginning of each EC steps randomly.

Additionally we want to point out that the factor $\beta$ in Eq.\ (\ref{eq:x}), which originates
from the rejection mechanism, is indeed the inverse of the thermodynamic temperature. If 
$\beta$ would be an arbitrary parameter, this would lead to a Fokker-Planck equation similar
to that of an ABP, but with a translational  diffusion constant $D$ and a rotational diffusion constant
$D_r$ corresponding to different temperatures (according to the Einstein relation), which would not make sense. 
Added to that, if $\beta$ was a parameter independent from the inverse temperature, one could simply
increase the parameter  $\beta \rightarrow \infty$ to  cause the translational diffusion term to vanish without 
affecting the temperature. 

To recapitulate: For the kEC algorithm to be exact for single ABPs, we require $x \rightarrow 0$ 
and, thus,  $\langle \Delta t_\text{EC} \rangle \rightarrow 0$. Since all physical properties of a system 
are fixed, the EC length is the only accessible parameter, we can modify, which effectively leads 
to the condition $\ell_\text{EC} / \sigma \rightarrow 0$. In reality, we can not choose arbitrary 
small displacement lengths without increasing the wall time of the simulation. Nonetheless reducing
$\ell_\text{EC}$ will always lead to better results, and we will see that $\ell_\text{EC} = 0.01\sigma$
leads to almost exact results, for all types of systems and parameters.
A choice  $\ell_\text{EC} \approx 1 \sigma$ is already small enough, that
deviations in the positional diffusion term of the Fokker-Planck equation should only influence
features of the distribution  $P(\vec{r}, \theta, t)$ 
below one particle diameter $\sigma$.

In order to support the claim, that the kEC algorithm becomes 
exact on the level of the probability distribution of a single ABP for 
$\ell_\text{EC} / \sigma \rightarrow 0$, we simulate 
single ABPs in a quadratic box of length $l_{\mathrm{box}} = 8\sigma$ 
with periodic boundary conditions and measure the one-particle 
probability density distribution $p_1(\vec{r},t)$ at $t = 1.5\tau$.
We choose $\mathrm{Pe} = 4 $ and compare kEC results with a
standard Brownian dynamics simulation, which was performed using LAMMPS. 
Fig.~\ref{fig:one_body_problem_probability_density_distribution}
 reveals that the
deviations between both simulation techniques 
(right column  of Fig.~\ref{fig:one_body_problem_probability_density_distribution}) 
gradually vanish with 
decreasing EC length $\ell_{\mathrm{EC}}$ 
from an  average absolute value of the relative error per grid element of 
$\langle \delta^{\ell_{\mathrm{EC}} = 1.00\sigma}_{\mathrm{1,kEC,LAMMPS}} \rangle \approx 26.4\%$ 
($\beta F_0 \ell_\text{EC} = 12$) to just $\langle \delta^{\ell_{\mathrm{EC}} =
0.01\sigma}_{\mathrm{1,kEC,LAMMPS}} \rangle \approx 2.3\%$ ($\beta F_0 \ell_\text{EC} = 0.12$).

These results are further supported by analyzing the mean square 
displacement of a single free ABP as a function of time for 
$\mathrm{Pe} = 50$ in Fig.~\ref{fig:one_body_problem_msd}, which shows 
again that errors decrease for  decreasing 
EC length $\ell_{\mathrm{EC}}$. This second comparison of a dynamic quantity
demonstrates convincingly that the kEC algorithm is not only able to 
produce almost exact results on short timescales but
conveniently also on medium and long timescales. 

\begin{figure}
        \includegraphics[width=1\linewidth]{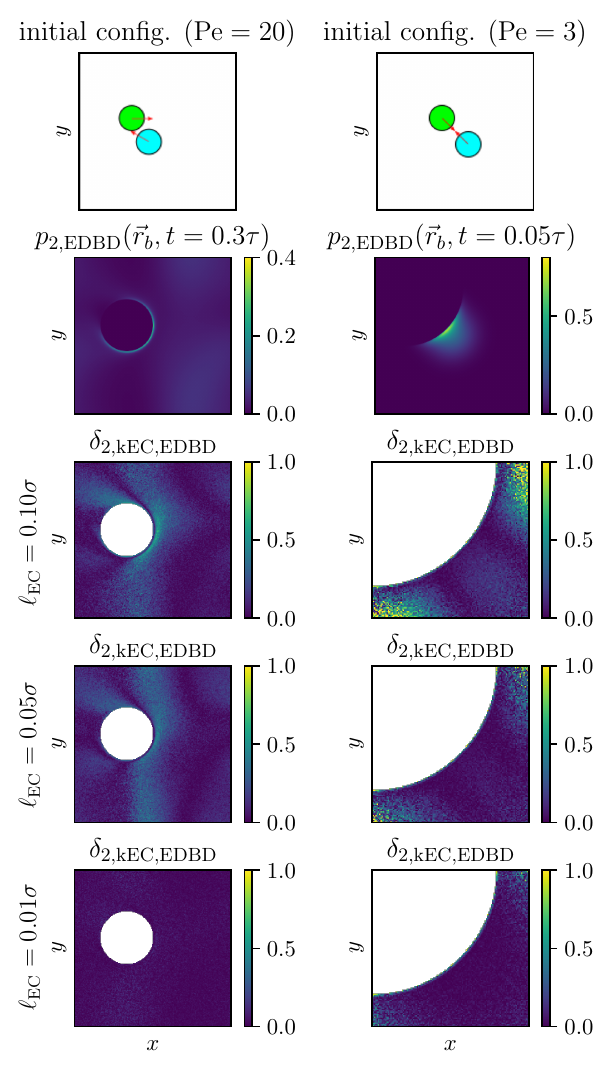}
    \caption{
    Probability density distributions $p_2(\vec{r}_b, t)$ of two hard ABPs 
    using the EDBD algorithm (see 
    second row, using $\tau_B = 0.01$) for two distinct initial 
    configurations (first row) and absolute 
    value of the relative error per grid element 
     $\delta_{\mathrm{2,kEC,EDBD}} \equiv \vert (p_{2,\mathrm{kEC}} 
    - p_{2,\mathrm{EDBD}}) / p_{2,\mathrm{EDBD}} \vert$ of the kEC algorithm. 
    We choose the coordinate system in a way that the position of the 
    green particle remains fixed and measure the position of the 
    blue particle. 
    The results demonstrate the kEC algorithm's ability to handle interactions 
    and collisions properly if $\ell_{\mathrm{EC}} / \sigma \rightarrow 0$.
    }
    \label{fig:two_body_problem}
\end{figure}

Even though the two-body problem is the most simple configuration, where
interactions between two particles are relevant, we are not able to analytically prove
the exactness of our algorithm on the level of a Fokker-Planck equation,
which takes collisions into account.
However, similar to the one-body problem, we can verify the validity of the 
kEC algorithm on the two-body level by comparing it to standard simulations of  two 
ABPs in a quadratic box of length $l_{\mathrm{box}} = 6\sigma$ with periodic 
boundary conditions. We choose the coordinate system in a way that the 
position of one particle (green) remains fixed and measure the probability 
density $p_2(\vec{r}_b,t)$ of the other 
particle (blue). Here, we compare our results with the EDBD algorithm 
previously used by Levis \textit{et al.} \cite{Levis2017}. 

In Fig.\ref{fig:two_body_problem}, we show  both initial configurations, 
the probability densities $p_2(\vec{r}_b,t)$ as well as heatmaps of the 
absolute value of the relative error per grid element. 
Regardless of the orientation of the active forces, the P\'{e}clet number as 
well as the time, the average deviation between both simulation techniques 
shrinks with decreasing EC length just like in the one-body problem. 
For a more detailed discussion we refer to the SM \cite{SM}, Sec.\ \ref{sec:validation_sm}).

\begin{figure}
   \includegraphics[width=1\linewidth]{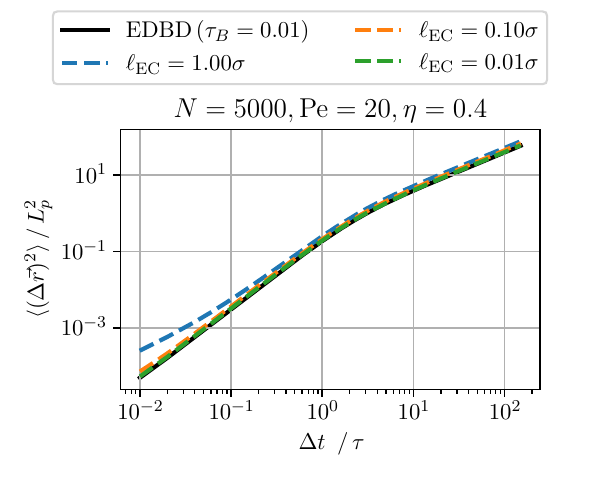}
   \includegraphics[width=1\linewidth]{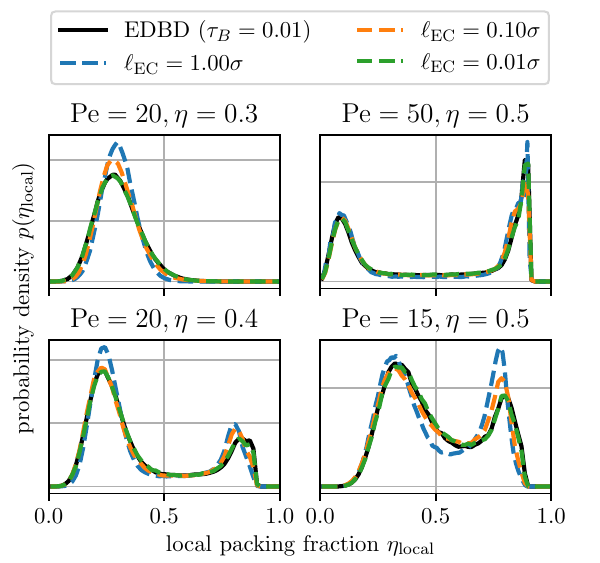}
    \caption{
    Comparison of a) the self diffusion and b) probability density of the 
    local packing fraction $p(\eta_{\mathrm{local}})$ for
    various configurations of hard ABPs using the EDBD algorithm 
    \cite{Levis2017} and the kEC algorithm.
    Using the parameters $\ell_{\mathrm{EC}} = 0.01\sigma$ and $\tau_B = 0.01$ 
    one obtains almost the exact same results.}
    \label{fig:local_packing_fraction}
\end{figure}

As mentioned before, we are not able to analytically
verify our algorithm as soon as collision
become relevant. However, we can prove the kEC algorithm to obey the $N$-body Fokker-Planck
equation in the dilute limit $\eta \rightarrow 0$ (for the derivation see SM \cite{SM}, 
Sec.\ \ref{sec:nbodyfokkerplanck_dilute}). Even though this limit itself is not very
interesting, one may argue that the algorithm does not become discontinuously incorrect, when 
the packing fraction is increased to finite values.
This theory is supported by the reliability of the kEC algorithm on the two-particle level.
Consequently we anticipate the algorithm to also work correctly for the general many-body 
case if the event chain length becomes sufficiently small because ECs involve only two 
particles in this limit.

Going beyond two particles, we can also test the validity 
of the many-particle kEC algorithm for
static and dynamic quantities. 
We simulate $N = 5000$ hard ABPs in a quadratic simulation volume of area $A$ and 
packing fraction~$\eta = N \pi \sigma^2 / (4A)$. Just like before the EDBD algorithm 
serves as the benchmark. We calculate the self-diffusion of ABPs and the probability
density~$p(\eta_{\mathrm{local}})$ of the local packing 
fraction~$\eta_{\mathrm{local}}$ (for further details see SM \cite{SM}, 
Sec.\ \ref{sec:calculation_distribution_local_packing_fraction}). 
The probability density $p(\eta_{\mathrm{local}})$ exhibits two peaks if MIPS occurs; positions, heights and shapes of the peaks provide a very sensitive test of the 
algorithm. The corresponding results of those measurements are shown in 
Fig.~\ref{fig:local_packing_fraction}. 
Comparing both algorithms reveals how relatively large EC lengths, e.g.~$\ell_{\mathrm{EC}} = 1.00\sigma$
($\beta F_0 \ell_\text{EC} = 45 - 150$), usually 
lead to sufficient approximations. In addition, reducing the EC length - 
as expected - improves the accuracy of the kEC algorithm to the point 
where, using~$\ell_{\mathrm{EC}} = 0.01\sigma$ ($\beta F_0 \ell_\text{EC} = 0.45 - 1.5$), both 
algorithms produce almost identical results. 
This holds for different values of the driving
force and global packing fraction, regardless of whether MIPS occurs or not. 

Concerning the individual simulation time of each particle: 
if the algorithm would distribute the mean 
time per EC-step falsely, one would definitely expect some deviations for 
almost every observable and
especially the ones that directly depend on time, such as self-diffusion. 
However, as just presented, 
the measurement of these very sensitive observables strongly indicates, 
that the limit~$\ell_\text{EC} / \sigma \rightarrow 0$ is 
conveniently also exact for the many-body problem. 
The reason for correctness in the limit of short ECs is probably a combination of a) an increased 
accuracy of single-particle
dynamics, b) a narrow time distribution (see SM \cite{SM}) and c) a decreasing 
mean number of  particle collisions per ECMC
move, effectively leading to two-body interactions, 
which our algorithm, as previously demonstrated, 
handles correctly.

\section{Motility-Induced Phase Separation}

\begin{figure*}
\includegraphics[width=1\linewidth]{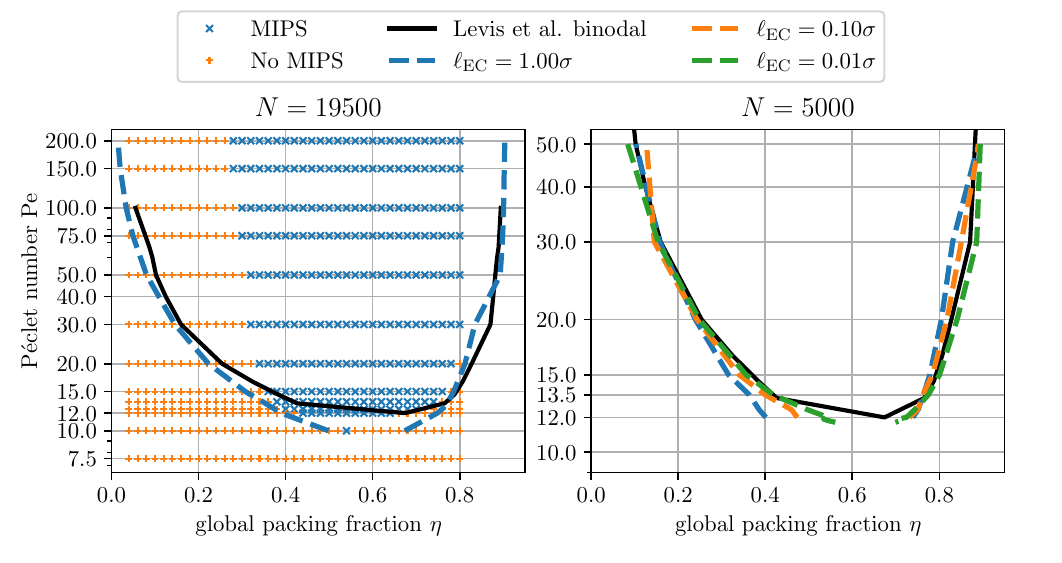}
\caption{
	Left: Phase diagram of active hard disks (kEC: $N = 19500$ 
	and area $A$ changing according to the packing 
	fraction~$\eta$) from kEC simulations
	using $\ell_{\mathrm{EC}} = 1.00\sigma$. 
	The kEC results
	for the binodal in general agree with EDBD simulation results 
	($N = 2000-4000$) 
	of Levis \textit{et al.} 
	\cite{Levis2017} except for some deviations around the critical point.
	Right: Phase diagram of active hard disks (kEC: $N = 5000$) 
	using different values for the EC length. In agreement 
	with the previous observations, reducing $\ell_{\mathrm{EC}}$ 
	leads to better results, especially around the critical 
	point. The remaining deviations are most likely an artifact of 
	different system sizes and ways of calculating the binodal.
	}
\label{fig:full_mips_phase_diagram}
\end{figure*}

There are numerous Brownian dynamics simulation studies of MIPS for active soft  repulsive disks  \cite{Fily2012,Bialke2013,Redner2013,Stenhammar2014,Martin-Roca2021} or spheres \cite{Stenhammar2014,Wysocki2014}, 
employing  Lennard-Jones interactions.
However, among all the numerical studies of active particles, 
results for genuine \emph{hard} disks 
are actually very rare \cite{Ni2013,Levis2017,DeMacedoBiniossek2018} and all based on the same 
EDBD algorithm developed originally for passive hard spheres \cite{scala2007event}.
As opposed to 
force-based Langevin or molecular dynamics techniques, the kEC algorithm is 
particularly suited to study hard particles. We demonstrate 
the potential of the novel kEC cluster algorithm 
to investigate such systems by calculating the MIPS
region in the phase diagram of active hard disks.
The phase diagram of active hard disks as a function of the 
packing fraction $\eta$ in a square system 
of area $A$ and the  P\'{e}clet number ${\rm Pe}$ as well as 
the extent of the MIPS region also provide 
another very sensitive test of our novel kEC algorithm.

Using $\ell_{\mathrm{EC}} = 1.00\sigma$, we can perform 
simulations of $N = 19500$ particles,
which are particle numbers 
that are hard to reach with the EDBD algorithm (Levis \textit{et al.}
only simulate $N = 2000-4000$
particles to investigate the phase diagram). 
Information about typical simulation times of our algorithm on an Intel Xeon W-2245 (8-Core, 
$3.9 \, \text{GHz}$) processor are listed in Tab. 
\ref{tab:typical_simulation_times}. For a more detailed discussion and a comparison with
the EDBD algorithm see Sec. \ref{sec:performance_comparison}.

\begin{table}[h]
    \centering
    \begin{tabular}{c|c|c}
        $\eta$      & Pe            & $\tau \, / \, \text{day}$ \\ \hline
        \, $0.3$ \, & \, $10$  \,     & $\sim 75000$ \\
        \, $0.4$ \, & \, $20$  \,     & $\sim  7000$ \\
        \, $0.5$ \, & \, $40$ \,      & $\sim  1000$ \\
        \, $0.6$ \, & \, $80$ \,      & $\sim   200$ 
    \end{tabular}
    \caption{To quantify the performance of our algorithm on an Intel Xeon W-2245 (8-Core, 
    $3.9 \, \text{GHz}$), we measure the number of persistence times $\tau$ we are able to simulate
    per day for different packing fractions $\eta$ and P\'{e}clet numbers Pe.}
    \label{tab:typical_simulation_times}
\end{table}

Starting from a homogeneous initial packing, we detect whether MIPS occurs (blue 
points in Fig.~\ref{fig:full_mips_phase_diagram}) and determine the 
corresponding local coexisting packing fractions (binodal, 
blue line) from measurements of  the probability
density~$p(\eta_{\mathrm{local}})$ of the local packing 
fraction~$\eta_{\mathrm{local}}$ as exemplified in  Fig.~\ref{fig:local_packing_fraction}.

Orange points within the binodal correspond to 
global packing fractions, where the homogeneous initial state remains metastable, 
such that the boundary between blue and orange points represents the 
spinodal line, i.e., the limit of stability of the homogeneous phase
(see SM \cite{SM}, Sec.\ \ref{sec:calculation_distribution_local_packing_fraction} for more details). 
The results, using $\ell_{\mathrm{EC}} = 1.00\sigma$, show good agreement 
with the active hard disk results of Ref.\ \citenum{Levis2017}, especially 
in the region of high activity. Noticeable deviations can only be found 
in close proximity of the critical point.

To further investigate this, we simulate the region around the critical point 
for a smaller system of $N = 5000$ particles and smaller values of $\ell_{\mathrm{EC}}$ 
(see right side of Fig.~\ref{fig:full_mips_phase_diagram}). 
In accordance with everything mentioned above,
we can correct the deviations between the EDBD algorithm 
and our kEC  algorithm by 
reducing $\ell_{\mathrm{EC}}$. Finally, the remaining differences, that are still present for 
$\ell_{\mathrm{EC}} = 0.01\sigma$, are almost certainly an artifact of 
different system sizes being analysed and more importantly, different ways of 
calculating the coexisting densities, relevant for determining 
the binodal line.

\section{Active particles with soft interactions}

\begin{figure}
    \includegraphics[width=1\linewidth]{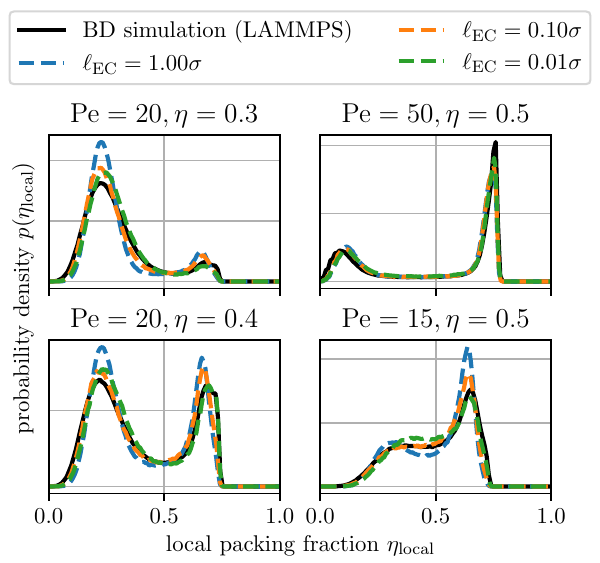}
    \caption{
    Comparison of the probability density of the local packing 
    fraction $p(\eta_{\mathrm{local}})$ of WCA particles  between the kEC algorithm using different values of the EC
    length and Brownian dynamics simulations conducted with LAMMPS. Similar to 
    hard-disk-systems a decrease in $\ell_{\mathrm{EC}}$ generally leads to better 
    agreement with the comparative results.}
    \label{fig:lennard_jones_local_packing_fraction}
\end{figure}

Up to this point, we restricted ourselves to the important case of 
steric interactions between hard active particles. Despite this focus, 
the ECMC algorithm can be extended to include soft interaction energies, 
by generalizing the notion of a rejection and introducing 
a ``virtual'' hard-disk radius \cite{Michel2014,kampmann2021}. 
The question arises, 
whether the kEC algorithm can be extended in a similar way.

The main problem when dealing with soft instead of steric 
interactions in the kEC algorithm 
is related to the mean time per EC step. 
In a many-body simulation of active hard disks, a particle 
behaves similar to a free particle, apart from the instantaneous collisions. 
We take advantage of this fact, by distributing the time per EC move to all 
particles contributing to an EC step, see Eq.\ (\ref{eq:tECN}). 
This is no longer true 
for soft interactions, because collisions are not instantaneous anymore. 
In principle,  we have to correct the mean time per EC 
step by taking the interaction forces into account, which is a 
highly non-trivial problem. 
In the following, we will investigate the kEC algorithm's ability 
to simulate active particles with soft interaction
energies without corrections of the mean time per EC as 
compared to Eq.\ (\ref{eq:tECN}). 

As an example system, we consider purely repulsive Weeks-Chandler-Andersen (WCA) particles  \cite{Weeks1971}
\begin{equation}
	V_{\mathrm{WCA}}(r) = 
	\begin{cases}
		4\epsilon 
		\left[
		\left( \frac{\sigma}{r} \right)^{12} - \left( \frac{\sigma}{r} \right)^6
		\right] + \epsilon
		, & r < \sqrt[6]{2}\sigma \\
		0,      & r \geq \sqrt[6]{2}\sigma.
	\end{cases} 
\end{equation}
In the following, we choose $\epsilon = 100k_BT$ and measure all 
lengths in units of $\sigma$ (i.e., use $\sigma = 1$ effectively) 
and do not apply any corrections to the mean time per EC step. 
We simulate $N = 5000$ 
particles using the kEC algorithm with generalized rejections for 
different values of the global packing fraction and the P\'{e}clet 
number. To compare and verify our results, we conduct Brownian dynamics simulations 
using LAMMPS.

Again, we perform  measurements of  the probability
density~$p(\eta_{\mathrm{local}})$ of the local packing 
fraction~$\eta_{\mathrm{local}}$ to  
detect whether MIPS occurs and determine the 
corresponding local coexisting packing fractions.
As Fig.\ \ref{fig:lennard_jones_local_packing_fraction} suggests, 
we obtain fair approximations of the probability densities of the local packing fraction using large EC lengths and very 
good results by reducing the EC length per step. 
As for hard disks with the EDBD algorithm, 
we obtain favorable agreement between the conventional Brownian dynamics algorithm and 
the kEC algorithm throughout the whole phase diagram for all global packing fractions 
and  P\'{e}clet  numbers. 
The only notable deviation of the kEC 
algorithm to the Brownian dynamics simulation is a slight off-set of the low-density, dilute phase 
peak if  MIPS occurs. This is 
probably a direct consequence of the necessity to alter the mean time
per EC step in many-body systems with soft
interaction energies. 

The example just mentioned proves the applicability of our 
algorithm to hard-disk-like soft interactions. This feature
makes our algorithm to stand out: while the EDBD algorithm is only 
applicable to steric interactions, Brownian dynamics simulations are limited to
simulate particles with soft interaction energies, while the
kEC algorithm can be used to investigate \emph{both} systems. However,
one should keep in mind that without any modifications to 
$\langle \Delta t_{\mathrm{EC}} \rangle$, we expect deviations to increase, 
e.g.\ if the range of the soft pair potential is increased or if its steepness is altered.

\section{Performance comparison}
\label{sec:performance_comparison}

\begin{figure}
	\includegraphics[width=1\linewidth]{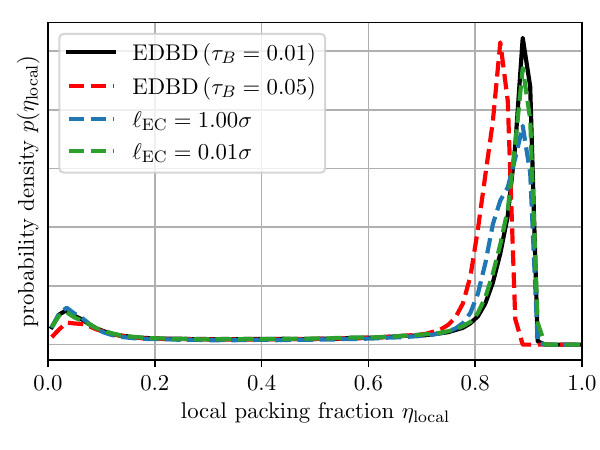}
	\caption{
		Probability density of the local packing fraction for a dense configuration 
		($\mathrm{Pe} = 70, \eta = 0.70, N = 5000$) of 
		active hard disk using the EDBD and kEC algorithm. Compared to $\tau_B = 0.01$, 
		which serves as a comparative result, $\ell_{\mathrm{EC}} = 0.01\sigma$ 
		also leads to exact results. Both $\tau_B = 0.05$ and $\ell_{\mathrm{EC}} 
		= 1.00\sigma$ provide good approximations.}
	\label{fig:local_packing_fraction_wall_time_comparison}
\end{figure}

\begin{figure}
		\includegraphics[width=1\linewidth]{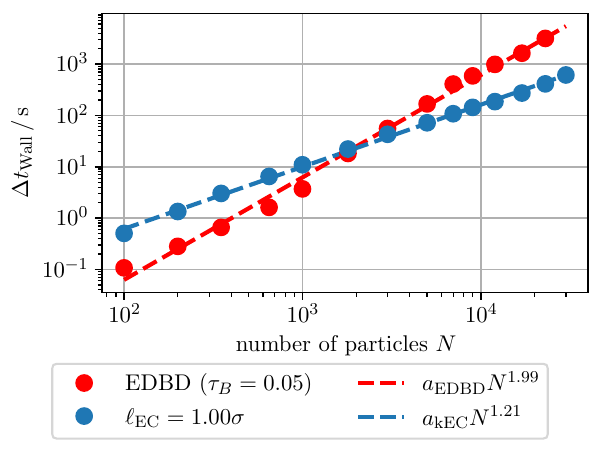}
		\includegraphics[width=1\linewidth]{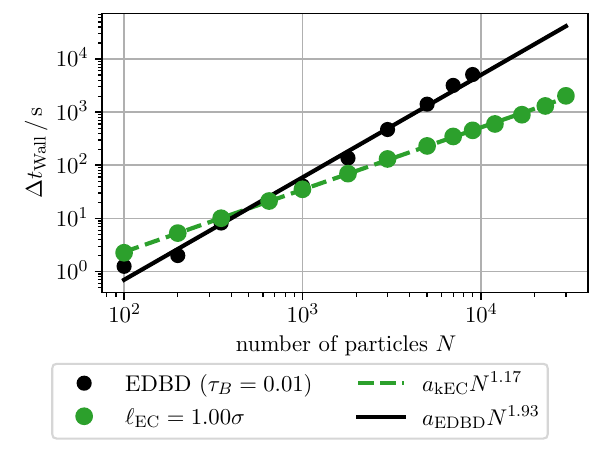}
	\caption{
			Upper: Performance comparison between the kEC algorithm~($\ell_{\mathrm{EC}} =
			1.00\sigma$) and the EDBD algorithm~($\tau_B = 0.05$) used by Levis {\it et 
			al.} \cite{Levis2017} for a dense configuration ($\mathrm{Pe} = 70, \eta = 0.70$), 
			with special focus on producing satisfactory
			approximations. 
			Lower: Performance comparison between the kEC algorithm~($\ell_{\mathrm{EC}} =
			0.01\sigma$) and the EDBD algorithm~($\tau_B = 0.01$) with special focus on 
			producing exact results. 	
			As the data suggests, the kEC algorithm is the superior algorithm regarding both
			comparisons when it comes to simulating	dense systems of active particles efficiently.}
	\label{fig:wall_time_comparison}
\end{figure}

Algorithms should be  both accurate and performant.
As mentioned before, the novel kEC algorithm is able to produce satisfactory 
approximations as well as exact results, depending on the EC length chosen. 
Because of the non-linear relationship
between the EC length~$\ell_{\mathrm{EC}}$ and the mean time per EC 
move~$\langle \Delta t_{\mathrm{EC}} \rangle$ (see Eq.\ (\ref{eq:tEC})) the total simulation time 
increases in the limit~$\ell_{\mathrm{EC}} / \sigma \rightarrow 0$. 
However, the EDBD algorithm, which serves as our benchmark for 
hard disks, suffers from a similar problem: to increase the accuracy 
one has to decrease the Brownian timestep $\tau_B$ \cite{Levis2017}, which again 
increases the required simulation time.
At this point it should be emphasized that both algorithms (or rather their frameworks)
use sophisticated neighbor lists to reduce the computational effort to detect collisions, 
e.g.\ the EDBD algorithm uses a cell list with a complexity of $\mathcal{O}(N)$.
However the EDBD algorithm has another shortcoming. One needs to perform Newtonian
dynamics simulations between each timestep and avoid possible overlaps by tracking
all possible collisions and resolve them in the same order as in a Newtonian dynamics 
simulation \cite{scala2007event}.

To demonstrate the performance of our algorithm, we measure the required wall time per 
persistence time $\tau$ of a dense system configuration ($\mathrm{Pe} = 70$ and $\eta = 0.70$, i.e., deep in the MIPS regime) 
for various system sizes and compare these results with the EDBD algorithm used 
by Levis \textit{et al.}~\cite{Levis2017}. For this purpose we use an Intel Xeon W-2245 (8-Core, 
$3.9 \, \text{GHz}$) processor, as before.
We proceed in the following way: first of all we focus on the question, which algorithm 
provides a satisfactory approximation more efficiently and afterwards take a closer look, 
which algorithm is more performant while producing accurate results.

Before we can compare both algorithms, we need to find suitable values 
of $\ell_\text{EC}$ and $\tau_B$.
For kEC simulations with small EC length 
$\ell_{\mathrm{EC}} = 0.01\sigma$ and EDBD simulations with 
small Brownian timestep $\tau_B = 0.01$, we essentially 
have agreement of the probability density of the local packing fraction as shown in Fig.~\ref{fig:local_packing_fraction_wall_time_comparison}. 
Therefore, this can be regarded as the quasi-exact result. 
We choose larger values 
$\ell_{\mathrm{EC}} = 1.00\sigma$ and $\tau_B = 0.05$ 
to be our simulation 
parameters for obtaining faster but approximative results 
also shown in 
Fig.~\ref{fig:local_packing_fraction_wall_time_comparison}. 
The only advantage of the EDBD simulation with $\tau_B = 0.05$
compared to kEC with 
$\ell_{\mathrm{EC}} = 1.00\sigma$ is an improved agreement of the shape of the  
of the probability density for high values of the local packing fraction, 
but with a non-negligible shift of the high-density peak. In general, the EDBD 
algorithm with large values of~$\tau_B$ does not pack the 
system as densely, 
as it is supposed to be. 
Apart from this our algorithm produces far 
better results in the low-density region as well as the correct position of the
high-density peak. 

Now we compare the wall time of the fast but approximative 
versions of the algorithms  with larger values 
$\ell_{\mathrm{EC}} = 1.00\sigma$ and $\tau_B = 0.05$
- see the upper part of Fig.~\ref{fig:wall_time_comparison}.
Both algorithms exhibit a power-law  dependence on the number of particles $N$, $\Delta t_{\mathrm{Wall}} = a N^k$. An appropriate fit gives exponents $k_{\mathrm{kEC}} \approx 1.21$ 
and~$k_{\mathrm{EDBD}} \approx 1.99$. The drastically reduced exponent  implies a substantial and qualitative 
performance gain of the kEC algorithm over EDBD simulation for large systems.  This performance gain 
is already  realized for relatively small systems with $N \gtrsim 1850$ particles.

We can also compare the wall time of the quasi-exact 
versions of the algorithms  with small values 
$\ell_{\mathrm{EC}} = 0.01\sigma$ and $\tau_B = 0.01$
- see the lower part of Fig.~\ref{fig:wall_time_comparison}.
Again we find a 
power-law  dependence on the number of particles $N$
with only slightly smaller exponents~$k_{\mathrm{kEC}} \approx 1.17$ and~$k_{\mathrm{EDBD}} \approx 1.93$. 
Therefore, the kEC algorithm 
also proves to be the qualitatively superior regarding the overall computation time when it comes to providing exact result
for large systems. 
For the chosen parameter values deep in the MIPS phase, the kEC algorithm is already faster 
 for $N \gtrsim 500$ particles.

Since both algorithms use neighbor list, as mentioned before, the process of tracking 
and resolving collisions takes (as expected) a toll on the computational complexity of 
the EDBD algorithm. Consequently regarding both comparisons, we expect the efficiency 
advantage of our algorithm to improve even further, the denser the entire system and 
the denser the  liquid phase gets if MIPS occurs, 
as the number of collisions per timestep 
will increase in dense regions.

\section{Giant system sizes}

\begin{figure}[h]
	\includegraphics[width=1\linewidth]{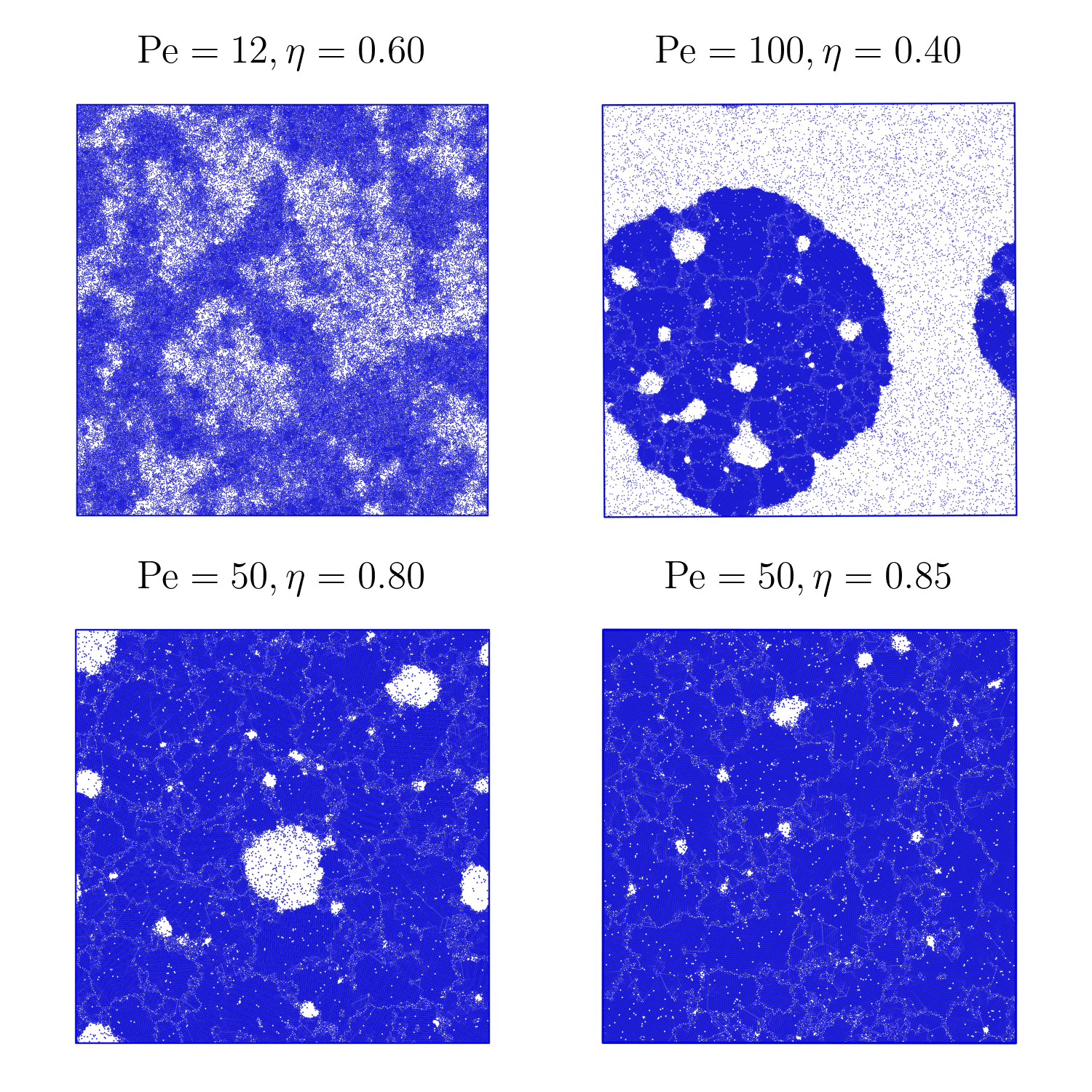}
    \caption{
            Exemplary snapshots of various configurations, each consisting of $N = 10^5$ particles.
            Upper left: System close to the critical point. Upper right: Formation of MIPS at high
            P\'{e}clet numbers. Lower row: High-density regime of the phase diagram. The emergence of
            small gas bubbles with various lifetimes within dense clusters is in agreement with 
            observations made in \cite{Caporusso2020}.}
	\label{fig:giant_systems}
\end{figure}

As demonstrated and quantified in the previous section, the performance of our algorithm is superior 
to the EDBD algorithm for  large and dense systems of ABPs. 
Therefore the kEC algorithm opens up the possibility to study the two most interesting regions of the
phase diagram, namely the critical point and the high-density regime, using large or even giant system
sizes to reduce finite-size effects. 
Fig.\ \ref{fig:giant_systems} shows
exemplary snapshots for a system containing 
$N = 10^5$  particles 
for various parameter values   
in different regions of the phase diagram.
All examples have reached their characteristic stationary 
state behavior using the kEC simulation technique. 

\section{Discussion}

We introduced a kinetic cluster MC algorithm for active systems, the kinetic event-chain (kEC)
algorithm to simulate active hard disks. 
To our knowledge, this is the first kinetic MC algorithm moving entire  
clusters of particles (in form of ECs) in order to achieve high numerical performance.
The basic idea is to distribute a mean time $\langle \Delta t_\text{EC} \rangle$, 
which is analytically calculated for a single particle, among all particles participating in a
cluster EC move under the action of the active forces and to use this mean time to rotate active
forces diffusive after each cluster move. This establishes a mapping onto ABPs.
Our rule to distribute the mean time yields constant currents along a moved EC, thus making the 
algorithm suitable to explore stationary states of this non-equilibrium system. 
The concept of cluster moves in a kinetic MC algorithm inevitably leads  to the concept of an  ``individual simulation time'' for each particle 
within this novel class of algorithms.
If we decrease activity to zero (and consider the limit of vanishing 
P\'{e}clet number $\mathrm{Pe} \to 0$), we essentially endow the ECMC algorithm for 
passive hard disks or spheres with an individual simulation time, which makes 
dynamical properties such as self diffusion behavior accessible to  ECMC simulations.
We can show that for the active (and passive) hard disk system the distribution of individual 
simulation times remains sufficiently narrow 
to warrant correct results. 

We analysed different observables, both of static and dynamic nature, for various one-, 
two- and many-body systems of active hard disks, as well as the phase diagram with special 
focus on the MIPS region. For all of these quantities, 
performance and accuracy of the kEC algorithm can be tuned 
via the EC length parameter $\ell_\text{EC}$: 
we are able to produce good approximations in fast simulations for $\ell_\text{EC} 
\approx 1\sigma$ and exact results in the limit $\ell_\text{EC} / \sigma \rightarrow 0$. We can rigorously prove
the algorithm to be exact in  this limit for the one-body problem and dilute $N$-body systems, as 
our algorithm correctly reproduces the corresponding Fokker-Planck equations.
We performed extensive numerical checks that the kEC algorithm is 
exact also on the two- and many-body level but are lacking 
an analytical proof, which we have to leave for future research. 

In addition, we successfully demonstrated the possibility to apply our method to soft particles,
by simulating purely repulsive active WCA particles, whose interaction forces only become relevant 
if the particles come very close. A generalization to arbitrary soft interactions will  require 
to adjust the mean time per EC move by taking the effect of interaction forces into account, which 
leads to non-trivial ``collisions'' of non-zero duration. Nevertheless, this 
makes our algorithm to stand out among other simulation techniques, as it would be possible to simulate 
steric and soft interactions, whereas other methods like an event-driven approach or a Brownian dynamics simulation 
are only able to handle one interaction type.

Finally, regarding active hard disks, a performance comparison revealed the kEC algorithm to be superior 
to the Event-Driven Brownian-Dynamics (EDBD) algorithm (at comparable accuracy). This performance gain is substantial 
for large systems as 
simulation times scale with significantly lower exponents with the number of
particles. This allows us to simulate large and dense systems of ABPs much more efficiently 
and may open the way for a more effective simulation of other active systems if the method can be generalized.
With the  kEC algorithm giant simulations  with $10^5$ active hard disks become possible on standard  desktop hardware. 

Since the ECMC method - the basis for our algorithm - is applicable across a wide range of physical problems,
we expect that with minor modifications the kEC algorithm can be extended to more complex active systems.
Potential applications include polydisperse mixtures of active and passive particles (relevant for studying phenomena 
such as active depletion forces or enhanced diffusion of passive tracers), the aforementioned
generalization to arbitrary soft interactions and the addition of external forces or other objects, such 
as polymers. Importantly, our method is not limited to two dimensions, as the core ideas can be readily extended to 
three-dimensional simulations.
Preliminary work on 
polydisperse mixtures of active and passive particles shows that 
additional challenges arise: the concept of an ``individual simulation time'' is harder to handle in 
this case, as active and  passive particles tend to develop distinct 
time distributions, which can cause difficulties, in particular for dynamic observables.

\section*{Author contributions} 
J.K. and T.A.K. conceived the project. N.S., T.S. and T.A.K designed the
simulations, which were performed and analysed by N.S. and T.S. The 
theoretical analysis was conducted N.S. and J.K. The manuscript was written
and revised by N.S. and J.K. with help of T.S. and T.A.K.

\section*{Data availability}
The simulation code for the kEC algorithm, as well as the analysis code, is openly 
available \cite{code_kEC}. All data generated with this code and presented in the Figures are available upon reasonable request from the authors.

\acknowledgments 
The authors are grateful to Demian Levis for providing the code of the Event-Driven
Brownian-Dynamics simulation framework. TAK acknowledges financial support by the Deutsche
Forschungsgemeinschaft (DFG) (Grant No.\ KA 4897/1–1).

\bibliography{paper}

\clearpage

\setcounter{section}{0}
\setcounter{subsection}{0}
\setcounter{subsubsection}{0}

\section*{Kinetic Event-Chain Algorithm for Active Matter: Supplemental Material}

\section{Details of the kEC algorithm}

\subsection{The algorithm}
\label{sec:algorithm}

The kEC algorithm is based on the ECMC algorithm for hard disks. 
Through factorization of the Metropolis filter each 
interaction can be handled entirely independently. Roughly speaking, 
each interaction gets a budget of a Boltzmann distributed energy 
that it uses to overcome energy barriers. 
The distance until the budget is empty sets the \emph{rejection distance} 
$d_{\text{rej}}$ for each interaction.
After all interactions are evaluated, the shortest distance
$\min_\text{interactions}(d_{\text{rej}})$ is moved and the EC is lifted to the
corresponding interaction partner, while the EC move
direction $\vec{e}_\text{EC}$ remains. An
EC is terminated when the sum of all displacements reach a pre-set
length $\ell_\text{EC}$.
In the presence of active forces, the corresponding linear one-particle
potentials can also trigger rejection, which results in lifting of the EC
direction to its reflection  with respect  to the
equipotential surface
($\vec{e}_\text{EC}' = \vec{e}_\text{EC} - 2 \mathbf{e}_i (\mathbf{e}_i \cdot
        \vec{e}_\text{EC})$), while  the moving particle remains the same. 

In a more structured way, an EC for active disks of diameters $\sigma_i$ is
constructed as follows:
\begin{itemize}
\item Choose random particle $i$ and direction $\vec{e}_\text{EC}$
  ($\lvert\vec{e}_\text{EC} \rvert = 1)$ and determine
  $d_{\text{rej}}$, next active
  particle $i_\text{next}$ and eventually new EC-direction
  $\vec{e}_\text{EC,next}$

\item  Initial values\\
$\phantom{uad}$ $d_{\text{rej}} = \ell_\text{EC}$\\
$\phantom{uad}$  $i_\text{next} = i$\\
$\phantom{uad}$  $\vec{e}_\text{EC,next} = \vec{e}_\text{EC}$

\item Hard Disks: \\
\textbf{for each} particle $j \neq i$ \\
$\phantom{uad}$ 	$\mathbf R = \mathbf r_j - \mathbf r_i$

$\phantom{uad}$ 	$\sigma_{ij} = 0.5(\sigma_i+\sigma_j)$

$\phantom{uad}$ 	$x = \mathbf R \cdot \vec{e}_\text{EC}$
	
$\phantom{uad}$	\textbf{if} $x>0$  {\scriptsize{(particles approach each other)}} 
	
$\phantom{qquad}$ 	$y = \sigma_{ij}^2 - \mathbf R^2 + x^2$
	
$\phantom{qquad}$ 	\textbf{if} $y>0$  {\scriptsize{(particles can hit each other)}} \\ 
$\phantom{uad}$	$\phantom{qquad}$  $d_{\text{rej},ij} =  x - \sqrt{y}$
	
$\phantom{uad}$	$\phantom{qquad}$ 	\textbf{if} $d_{\text{rej},ij} < d_{\text{rej}}$  {\scriptsize{(event before current one)}} 

$\phantom{qquad}$	$\phantom{qquad}$ 	$d_{\text{rej}}=d_{\text{rej},ij}$
		
$\phantom{qquad}$	$\phantom{qquad}$ 	$i_\text{next} = j$
		
$\phantom{qquad}$	$\phantom{qquad}$ 	$\vec{e}_\text{EC,next}
= \vec{e}_\text{EC}$
		
\item Active Force (\textbf{if} $F_{0,i} \neq 0$):
	
	$x = F_{0,i} (\mathbf{e}_i \cdot \vec{e}_\text{EC})$
	
	\textbf{if} $x<0$  {\scriptsize{(particle moves against the force)}} 
	
$\phantom{uad}$	$y = \text{rand}[0\ldots 1]$

$\phantom{uad}$	$d_{\text{rej},\text{active}} = {\text{ln} y}/{ \beta x}$
	
$\phantom{uad}$	\textbf{if} $d_{\text{rej},\text{active}} < d_{\text{rej}}$ {\scriptsize{(event before current one)}} 

	$\phantom{qquad}$	$d_{\text{rej}}=d_{\text{rej},\text{active}}$
		
	$\phantom{qquad}$	$i_\text{next} = i$
		
	$\phantom{qquad}$	$\vec{e}_\text{EC,next} =
        \vec{e}_\text{EC} - 2 \mathbf{e}_i (\mathbf{e}_i \cdot
        \vec{e}_\text{EC})$

\item Execute:

$\phantom{uad}$ $\mathbf r_i \mathrel{+}= d_{\text{rej}} \vec{e}_\text{EC}$

$\phantom{uad}$ $d_i \mathrel{+}=  d_{\text{rej}}$ 

$\phantom{uad}$  $\ell_\text{EC} = \ell_\text{EC} - d_{\text{rej}}$
	
$\phantom{uad}$  $i=i_\text{next} $

$\phantom{uad}$ $\vec{e}_\text{EC}= \vec{e}_\text{EC,next}$

\item \textbf{Repeat until} $\ell_\text{EC} =0$

\item Rotate active forces:

\textbf{for each} particle $i$

$\phantom{uad}$ $\langle \Delta t_{EC} \rangle_i = \frac{d_i}{\ell_{EC}} \langle \Delta t_\text{EC} \rangle$

$\phantom{uad}$ $\theta_i \mathrel{+}=  \text{ Gauss}(0, \sqrt{ 2 D_r  \langle t_{EC} \rangle_i} ) $

$\phantom{uad}$		$d_i=0$

\item Loop back to beginning
		
\end{itemize}
Here $\text{rand} (\dots)$ is a uniformly  distributed and 
$\text{Gauss} (\dots)$ is a Gaussian distributed random number.


\subsection{Mean ECMC move vector \texorpdfstring{$\langle \Delta \vec{r}_\text{EC} \rangle$ for a single particle}{}}
\label{sec:meantEC}

We consider a single particle under a constant force $F_0 \vec{e}$ and 
calculate its mean move vector $\langle \Delta \vec{r}_\text{EC} \rangle$
during one algorithmic ECMC step. 
The ECMC algorithm draws a  random direction $\vec{e}_\text{EC}$
and moves the particle by the EC length $\ell_\text{EC}$. 
We introduce the angle $\phi$ between the unit propulsion direction vector $ \vec{e}$ 
and  the EC unit vector $\vec{e}_\text{EC}$ 
by $\cos\phi \equiv \vec{e}_\text{EC}\cdot \vec{e}$.  The 
mean move vector   
$\langle \Delta \vec{r}_\text{EC}\rangle$ will be averaged 
over all angles $\phi$, i.e., over all EC directions because EC directions 
are chosen randomly in the algorithm and we are interested in 
the mean algorithmic displacement over many ECMC moves.
Three cases have to be considered in the
calculation of 
$\langle \Delta \vec{r}_\text{EC} \rangle$ 
depending on $\phi$ and $\ell_\text{EC}$, 
and we have to properly average over these cases and all angles 
$\phi$. 

(i)  If $\cos\phi \ge 0$, the move is energetically downhill. 
Then the particle is displaced by the full EC length $\ell_\text{EC}$ resulting 
in a mean displacement 
\begin{equation}
  \langle \Delta \vec{r}_\text{EC} \rangle_{\cos\phi \ge 0}
  = \ell_\text{EC}  \vec{e}_\text{EC}
\end{equation}  
for given $\phi$.
Averaging  over all angles $\phi \in [-\pi/2,\pi/2]$ with
$\cos\phi \ge 0$, we  obtain
\begin{align}
  \langle \Delta \vec{r}_\text{EC}  \rangle_{\ge }
  &= \frac{1}{\pi} \int_{-\pi/2}^{\pi/2} \mathrm{d}\phi
    \langle \Delta \vec{r}_\text{EC} \rangle_{\cos\phi \ge 0}
    \nonumber\\
  &= \frac{1}{\pi}\ell_\text{EC} \int_{-\pi/2}^{\pi/2}  \mathrm{d}\phi \cos\phi {\vec{e}}
    = \frac{2}{\pi}\ell_\text{EC}{\vec{e}}.
\end{align}

(ii)  
If $\cos\phi < 0$, the move is energetically uphill and a
``usable'' energy $\Delta U>0$ for uphill motion is drawn from an 
exponential Boltzmann distribution
$p(\Delta U) \sim \exp(-\beta \Delta U)$ ($\beta= 1/ k_\text{B} T$) 
and a rejection distance $d_\text{rej}$ is 
determined from $\Delta U =  F_0 d_\text{rej}|\cos\phi|$. This is 
equivalent to drawing the rejection distance from an exponential 
distribution
\begin{equation}
	p_d(d_\text{rej}) =  \beta F_0 |\cos\phi|
	e^{-\beta F_0 d_\text{rej}|\cos\phi| }.
	\label{eq:rejection_distance_distribution}
\end{equation}
 
Now, two subcases arise:
\begin{itemize}
	\item[(iia)] 
	$d_\text{rej} \ge  \ell_\text{EC}$, where the particle is moved 
	by the full event chain length,
  $\Delta \vec{r} = \ell_\text{EC}\vec{e}_\text{EC}$.
  This happens with probability
  \begin{equation}
    p_>  = \int_{\ell_\text{EC}}^\infty dx  p_d(x) =    
 e^{-\beta F_0 \ell_\text{EC} |\cos\phi| }.
\end{equation}
This subcase will give the dominating contribution in the limit 
$\beta F_0 \ell_\text{EC}\ll 1$ of small EC lengths or weak propulsion forces.

   \item[(iib)]  $d_\text{rej} <  \ell_\text{EC}$, where
  the particle is only  moved by  $\Delta \vec{r} =
 d_\text{rej}\vec{e}_\text{EC}$. Then, 
   the EC direction is lifted 
   $\vec{e}_\text{EC} \to \vec{e}_\text{EC}'$
 by reflecting with respect to the
 equipotential surface,
 $\vec{e}_\text{EC}' = \vec{e}_\text{EC} - 2 \cos\phi \vec{e}$.
 In total, this results
 in $\Delta \vec{r} =d_\text{rej} \vec{e}_\text{EC}+
 (\ell_\text{EC}-d_\text{rej}) \vec{e}_\text{EC}'$.
 This displacement is realized with probability $p_d(d_\text{rej})$.
 This subcase will give the dominating contribution in the limit 
$\beta F_0 \ell_\text{EC}\gg 1$ of large EC lengths or strong propulsion forces. 
 \end{itemize}

  Averaging over $p_d(d_\text{rej})$ we obtain for 
 the  mean vectorial displacement 
\begin{align}
 & \langle \Delta \vec{r}_\text{EC} \rangle_{\cos\phi <0}\nonumber\\
  &= p_> \ell_\text{EC}  \vec{e}_\text{EC} 
   +\int_0^{\ell_\text{EC}} dx p_d(x) \left[x \vec{e}_\text{EC}+
    (\ell_\text{EC}-x) \vec{e}_\text{EC}'\right]
    \nonumber\\
  &= p_> \ell_\text{EC}  \vec{e}_\text{EC}\nonumber\\
  &~~  +\int_0^{\ell_\text{EC}} dx p_d(x) \left[\ell_\text{EC}\vec{e}_\text{EC}
    -2\cos\phi  (\ell_\text{EC}-x)\vec{e}\right]
    \nonumber\\
  &= \ell_\text{EC}  \vec{e}_\text{EC}\nonumber\\
  & ~~-2\cos\phi \vec{e}
    \int_0^{\ell_\text{EC}} dx p_d(x) (\ell_\text{EC}-x)
    \nonumber\\
  &= \ell_\text{EC}  \vec{e}_\text{EC}\nonumber\\
   &~~-\vec{e}
    \frac{2}{\beta F_0} \left(1-\beta F_0 \ell_\text{EC} |\cos\phi|
     - e^{-\beta F_0 \ell_\text{EC} |\cos\phi|} \right)      
\end{align}  
for given $\phi$.
Averaging over the all angles $\phi \in [\pi/2,3\pi/2]$ with
$\cos\phi < 0$, we obtain
\begin{align}
 & \langle \Delta \vec{r}_\text{EC}  \rangle_{<}\nonumber\\
  &= -\frac{1}{\pi}  \ell_\text{EC} \int_{-\pi/2}^{\pi/2}  \mathrm{d}\phi
    \cos\phi {\vec{e}}\nonumber\\
  &~~  - \vec{e}\frac{2}{\beta F_0} \frac{1}{\pi} \int_{-\pi/2}^{\pi/2} \mathrm{d}\phi
     \left(1-\beta F_0 \ell_\text{EC} |\cos\phi|
    - e^{-\beta F_0 \ell_\text{EC} |\cos\phi|} \right)
    \nonumber\\
  &= \frac{2}{\pi} \ell_\text{EC}\vec{e}
    - \frac{2}{\beta F_0}  \frac{1}{\pi} \int_{-\pi/2}^{\pi/2}  \mathrm{d}\phi
    \left( 1- e^{-\beta F_0 \ell_\text{EC} |\cos\phi|} \right)\vec{e}
\end{align}

\begin{figure}
\begin{center}
 \includegraphics[width=0.45\textwidth]{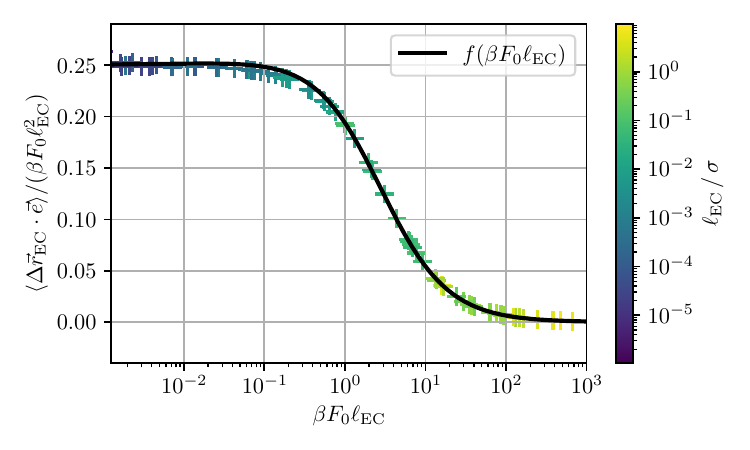}
\caption{
	Mean ECMC move lengths in force direction $\langle \Delta \vec{r}_\text{EC} \cdot \vec{e} \rangle$
	in units of $\beta F_0 \ell_\text{EC}^2$ as a function of $x=\beta F_0 \ell_\text{EC}$. The kEC
	simulation data (crosses) for different EC lengths  $\ell_\text{EC}$ (color-coded) collapse onto the analytically calculated scaling 
	function $f(x)$ from Eq.\ (\ref{eq:fx}) in the main text and 
	Eqs.\ (\ref{eq:exact}) and (\ref{eq:project})
	(black line, to reduce the computational effort the fit function
	(\ref{eq:fitf}) is plotted).
}
\label{fig:ec_time}
\end{center}
\end{figure}

Averaging over the two cases
$\cos\phi \ge 0$ and $\cos\phi <0$, we finally obtain
\begin{align}
  \langle \Delta \vec{r}_\text{EC}  \rangle
  &= \frac{1}{2}\left( 
         \langle \Delta \vec{r}_\text{EC} \rangle_{\ge } +
         \langle \Delta \vec{r}_\text{EC} \rangle_{< }
           \right)
    \nonumber\\
  \frac{\langle \Delta \vec{r}_\text{EC}  \rangle}{\ell_\text{EC}}
  &=  \frac{2}{\pi} \vec{e}
    \nonumber\\
  &  ~~- \frac{1}{\beta F_0 \ell_\text{EC}}
    \frac{1}{\pi} \int_{-\pi/2}^{\pi/2} \mathrm{d}\phi
    \left( 1- e^{-\beta F_0 \ell_\text{EC} |\cos\phi|} \right)\vec{e}
    \nonumber\\
  &= x f(x)\vec{e}
  \label{eq:exact}
  \end{align}
 with 
 \begin{equation}
       x\equiv \beta F_0 \ell_\text{EC} = v_0 \ell_\text{EC} / D =
         3\mathrm{Pe} \ell_{\mathrm{EC}} / \sigma 
       \label{sup-eq:x}
 \end{equation}
and a scaling function 
\begin{equation}
	f(x) = \frac{2}{\pi} \frac{1}{x} -
	\frac{1}{x^2} \frac{1}{\pi} \int_{-\pi/2}^{\pi/2} d\phi
	(1- e^{-x\cos\phi})
	\label{sup-eq:fx}
\end{equation}
as in  Eq.\ (\ref{eq:fx}) in the main text. 
It is also apparent that the projected mean EC move length is simply related via 
\begin{equation}
     \langle \Delta \vec{r}_\text{EC} \cdot {\vec{e}} \rangle = 
     \langle \Delta \vec{r}_\text{EC}  \rangle\cdot{\vec{e}}.
      \label{eq:project}
\end{equation}
The function $f(x)$ has the limiting behaviors $f(x) \approx 1/4 - 2x/9\pi$ for $x\ll 1$
(in the limit $\beta F_0 \ell_\text{EC}\ll 1$ of small EC lengths or weak propulsion forces)
and $f(x) \approx 2/\pi x - 1/x^2$ 
for $x\gg 1$ (in the limit  $\beta F_0 \ell_\text{EC} \gg 1$ of large EC lengths or strong propulsion forces). 
Figure \ref{fig:ec_time} shows perfect agreement of 
the analytically calculated scaling function $f(x)$  with the kEC simulation.

The limiting cases $x\ll 1$ and $x\gg 1$ differ in their behaviour under an 
opposing force ($\cos\phi <0$). For $x\ll 1$,  
the particle moves with probability $p_>$ the
whole EC length against a weak opposing force resulting in 
$\langle \Delta \vec{r}_\text{EC} \rangle_{\cos\phi <0}\approx  p_> \ell_\text{EC}  \vec{e}_\text{EC} = e^{-\beta F_0 \ell_\text{EC} |\cos\phi|}\ell_\text{EC}  \vec{e}_\text{EC}$ (corrections from 
lifting and reflection are of order $F_0^2$), 
while $\langle \Delta \vec{r}_\text{EC} \rangle_{\cos\phi \ge 0} = \ell_\text{EC}  \vec{e}_\text{EC}$ for a move in the opposite direction $-\vec{e}_\text{EC}$. 
This leads to exactly the same behaviour as in a local MC algorithm 
for a local MC move of length $\ell_\text{EC}$ in directions $\pm \vec{e}_\text{EC}$ 
under the standard Metropolis rule fulfilling detailed balance, thus 
establishing equivalence to local Metropolis MC in the limit of small EC lengths $\ell_\text{EC}$. After expanding in $x\ll 1$ and 
averaging over all directions 
this gives $\langle \Delta \vec{r}_\text{EC} \rangle  \approx 
(1/4)\beta F_0 \ell_\text{EC}^2 \vec{e}$.

 For $x\gg 1$, on the other hand, the particle is lifted after a short mean 
 distance $1/\beta F_0|\cos\phi|$ in direction against a strong opposing force and reflected. 
Essentially, this results in a move in the reflected direction over the whole EC length, $\langle \Delta \vec{r}_\text{EC} \rangle_{\cos\phi <0}\approx 
 \ell_\text{EC}  \vec{e}_\text{EC}+2|\cos\phi|\ell_\text{EC}{\vec{e}}$. After averaging 
 over all directions this gives 
 $\langle \Delta \vec{r}_\text{EC} \rangle_{<}  \approx -(2/\pi)\ell_\text{EC}\vec{e} + (4/\pi)\ell_\text{EC}\vec{e}$ 
 resulting in  $\langle \Delta \vec{r}_\text{EC} \rangle  \approx (2/\pi)\ell_\text{EC} \vec{e}$.

In practice, the following fit function for $f(x)$ is useful 
to interpolate between both limiting cases 
(see also Fig.\ \ref{fig:ec_time}):
\begin{equation}
  f(x) =   \frac{1}{ax + b\sqrt{x} + c}
  \label{eq:fitf}
\end{equation}
with $a= \pi/2$, $b=-0.42$, and $c=4$, which is constructed to satisfy the limits $f(x) \approx 1/4$ for $x\approx 0 $ and 
$f(x) \approx 2/\pi x$ 
for $x\to \infty$, so that $b$ is the only free parameter. This fit function can be used to speed up calculations in production runs.

\subsection{Mean square and mean cubic ECMC move lengths for a single particle}
\label{sec: higher_moments}

Following similar steps, we can also calculate the mean square and mean cubic 
move lengths $\langle (\Delta \vec{r}_\text{EC}\cdot\vec{e})^2 \rangle$,
$\langle (\Delta \vec{r}_\text{EC}\cdot\vec{e})^3 \rangle$ and $\langle 
(\Delta \vec{r}_\text{EC} )^2 \rangle$ during one algorithmic 
ECMC step. In the following we use $x \equiv \beta F_0 \ell_{\mathrm{EC}}$ from 
Eq.\ (\ref{sup-eq:x}).

\subsubsection{The values of \texorpdfstring{$\langle (\Delta \vec{r}_\mathrm{EC} \cdot \vec{e})^2 \rangle$}{}}

If $\cos \phi \ge 0$:
\begin{align}
	\frac{\langle (\Delta \vec{r}_{\text{EC}} \cdot \vec{e})^2
  	\rangle_{\cos \phi \ge 0}}{\ell_{\text{EC}}^2} &= \cos^2 \phi 
  	 \\
\frac{\langle (\Delta \vec{r}_{\text{EC}} \cdot \vec{e})^2 
	\rangle_{\ge}}{\ell_{\text{EC}}^2} &= 
	\frac{1}{2}.
\end{align}
If $\cos \phi < 0$:
\begin{align}
\frac{\langle (\Delta \vec{r}_{\text{EC}} \cdot \vec{e})^2
	\rangle_{cos \phi < 0}}{\ell_{\text{EC}}^2} 
	&=(\cos \phi)^2 - \frac{4}{x^2} \bigg(
	 x|\cos \phi| - 2 \nonumber\\
	&~~~+ \big(x |\cos \phi| + 2\big)
	e^{-x |\cos \phi|} \bigg) 
	\\
\frac{\langle (\Delta \vec{r}_{\text{EC}} \cdot \vec{e})^2
	\rangle_{<}}{\ell_{\text{EC}}^2} &= 
	\frac{1}{2} - \frac{4}{\pi x^2} \int_{-{\pi}/{2}}^{{\pi}/{2}} \mathrm{d}\phi \,
	\bigg( x |\cos \phi| - 2 \nonumber\\ 
	&~~~+\big( x | \cos \phi| + 2\big) 
	e^{- x |\cos \phi|} \bigg).
\end{align}
The final result after averaging over the two cases
$\cos\phi \ge 0$ and $\cos\phi <0$ is
\begin{align}
	\frac{\langle (\Delta \vec{r}_{\text{EC}} \cdot \vec{e})^2
	\rangle}{\ell_{\text{EC}}^2} 
	&=\frac{1}{2} - \frac{2}{\pi x^2} \int_{-{\pi}/{2}}^{{\pi}/{2}} \mathrm{d}\phi \,
	\bigg( x |\cos \phi| - 2 \nonumber\\ 
	&+\big( x |\cos \phi | + 2\big) 
	e^{ - x | \cos \phi |} \bigg).
\end{align}

\subsubsection{Derivation of \texorpdfstring{$\langle (\Delta \vec{r}_\mathrm{EC} )^2 \rangle$}{}}

If $\cos \phi \ge 0$:
\begin{align}
	\frac{\langle (\Delta \vec{r}_{\text{EC}})^2
	\rangle_{\cos \phi \ge 0}}{\ell_{\text{EC}}^2} &= 1\\
	\frac{\langle (\Delta \vec{r}_{\text{EC}})^2 
	\rangle_{\ge}}{\ell_{\text{EC}}^2} &= 
	1.
	\label{eq:delta_r2>}
\end{align}
If $\cos \phi < 0$:
\begin{align}
	\frac{\langle (\Delta \vec{r}_{\text{EC}} )^2
	\rangle_{\cos \phi < 0}}{\ell_{\text{EC}}^2} 
	&= 1 - \frac{4}{x^2} \bigg(
	 x | \cos \phi | - 2 \nonumber\\
	&~~~~~~+ \big(x |\cos \phi | + 2\big)e^{-x|\cos \phi |} \bigg) 
	\\
	\frac{\langle (\Delta \vec{r}_{\text{EC}} )^2
	\rangle_{<}}{\ell_{\text{EC}}^2} 
	&=1 - \frac{4}{\pi x^2} \int_{-{\pi}/{2}}^{{\pi}/{2}} 
	\mathrm{d}\phi \,
	\bigg( x |\cos \phi| - 2 \nonumber\\ 
	&~~~~~~+\big( x |\cos \phi | + 2\big) 
	e^{ - x |\cos \phi | } \bigg).
\end{align}
The final result after averaging over the two cases
$\cos\phi \ge 0$ and $\cos\phi <0$ is
\begin{align}
\frac{\langle (\Delta \vec{r}_{\text{EC}})^2
	\rangle}{\ell_{\text{EC}}^2} 
	&= 1 - \frac{2}{\pi x^2} \int_{-{\pi}/{2}}^{{\pi}/{2}} 
	\mathrm{d}\phi \,
	\bigg( x |\cos \phi |- 2 
	 \nonumber\\ 
	 &~~~~~~+\big( x |\cos \phi | + 2\big) 
	 e^{ - x |\cos \phi | } \bigg)
	 \label{eq:delta_r2}.
\end{align}
We find ${\langle (\Delta \vec{r}_{\text{EC}})^2
	\rangle}/{\ell_{\text{EC}}^2} \approx 1$ both in the limits $x\gg 1 $ and $x\ll 1$,
	see also Fig.\ \ref{fig:validation_kEC_algorithm_eom} (bottom, orange line).

\subsubsection{Derivation of \texorpdfstring{$\langle (\Delta \vec{r}_\mathrm{EC} \cdot \vec{e})^3 \rangle$}{}}

If $\cos \phi \ge 0$:
\begin{align}
    \frac{\langle ( \Delta \vec{r}_\text{EC} \cdot \vec{e} )^3 \rangle_{\cos \ge 0}}{\ell_{\text{EC}}^3} 
    = \cos^3 \phi \\
    \frac{\langle ( \Delta \vec{r}_\text{EC} \cdot \vec{e} )^3 \rangle_{\ge}}{\ell_{\text{EC}}^3} 
    = \frac{4}{3\pi}.
\end{align}
If $\cos \phi < 0$:
\begin{align}
    \frac{\langle ( \Delta \vec{r}_\text{EC} \cdot \vec{e} )^3 \rangle_{\cos < 0}}{\ell_{\text{EC}}^3} 
    = \cos^3 \phi (A + B - 1) \\
    \frac{\langle ( \Delta \vec{r}_\text{EC} \cdot \vec{e} )^3 \rangle_{<}}{\ell_{\text{EC}}^3} 
    = \frac{1}{\pi} \int_{\pi / 2}^{3 \pi / 2} \mathrm{d}\phi \, \frac{\langle ( \Delta \vec{r}_\text{EC} 
    \cdot \vec{e} )^3 \rangle_{\cos < 0}}{\ell_{\text{EC}}^3}
\end{align}
using
\begin{align}
    A = &\frac{1}{x^3 \vert \cos \phi \vert^3} \Big( 6 x^2 \vert \cos \phi \vert^2 - 24 x \vert \cos \phi \vert 
    + 48 \Big) \\
    B = &- \frac{\exp(-x \vert \cos \phi \vert)}{x^3 \vert \cos \phi \vert^3} 
    \Big(6 x^2 \vert \cos \phi \vert^2 + 24 x \vert \cos \phi \vert + 48 \Big).
\end{align}
The final result after averaging over the two cases
\begin{align}
    \frac{\langle ( \Delta \vec{r}_\text{EC} \cdot \vec{e} )^3 \rangle}{\ell^3_\text{EC}}
    = \frac{1}{2 \pi} \int_{\pi / 2}^{3 \pi / 2} \mathrm{d} \phi \,
    \cos^3 \phi (A + B - 2).
\end{align}

\section{Validation of the kEC algorithm}
\label{sec:validation_sm}

\subsection{One-body problem: the kEC algorithm becomes exact for small $\ell_{\mathrm{EC}}$}

\begin{figure}
	\includegraphics[width=1\linewidth]{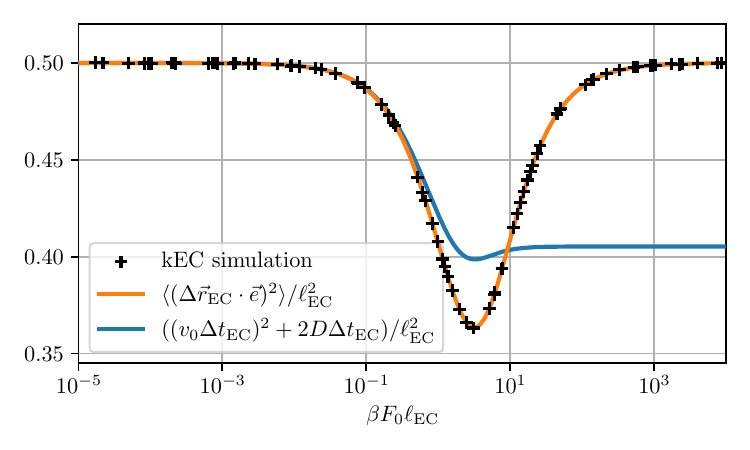}
	\includegraphics[width=1\linewidth]{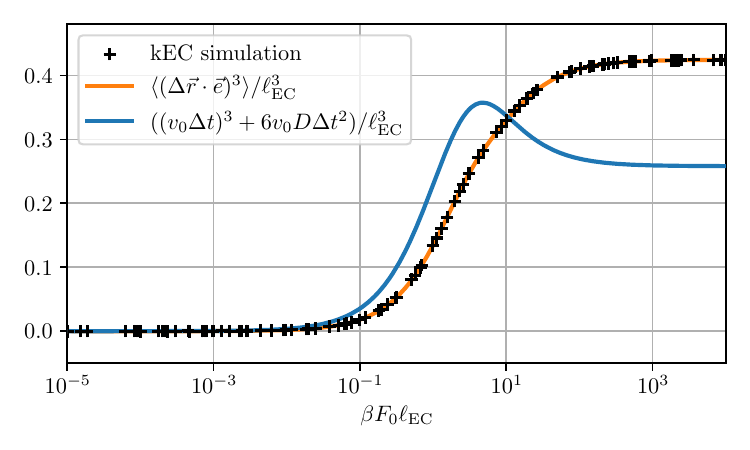}
    \includegraphics[width=1\linewidth]{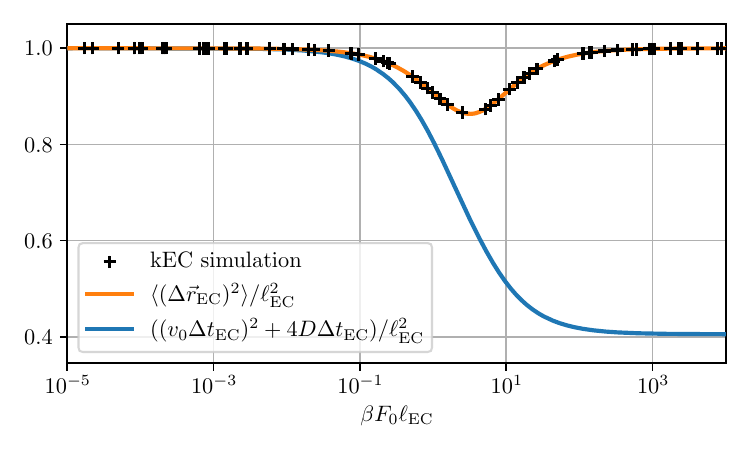}
    \caption{
        Higher moments of the ECMC move lengths (left-hand sides of 
	    Eqs.\ \ref{eq:eom1}\,-\,\ref{eq:eom3}) normalized to the $n$-th power of the EC length $\ell_{\mathrm{EC}}$ as a function of $x=\beta F_0 \ell_\text{EC}$. 
	    The kEC simulation data (black crosses) agree with analytical calculations (orange lines). 
		Left and right hand sides of 
	    Eqs.\ \ref{eq:eom1}\,-\,\ref{eq:eom3} evaluated with 
	    $\Delta t = \langle \Delta t_\text{EC} \rangle $  (blue lines) 
	    agree with the orange lines in the 
		limit $\beta F_0 \ell_{\mathrm{EC}} \rightarrow 0$, indicating
		that the limit  $v_0 \ell_{\mathrm{EC}} / D 
		\rightarrow 0$ reproduces correct physical behaviour on a single step level.
		}
	\label{fig:validation_kEC_algorithm_eom}
\end{figure}

We assign a kEC simulation time $\langle \Delta t_\text{EC} \rangle$ to each particle 
according to the relation 
$\langle \Delta \vec{r} \cdot \vec{e} \rangle= v_0 \Delta t$ for the \emph{first} moment of 
the displacement in force direction, which 
holds exactly for single free ABPs by requiring that 
$\langle \Delta \vec{r}_\text{EC} \cdot \vec{e} \rangle= v_0 \langle \Delta t_\text{EC} \rangle$
(see main text and preceding section). 
This procedure works equally well for all EC lengths $\ell_\text{EC}$
as demonstrated in Fig.\ \ref{fig:ec_time} and, thus, does not single out an optimal EC length. 
An optimal EC length could emerge if we also 
consider \emph{higher} moments $\langle (\Delta \vec{r} \cdot \vec{e})^n \rangle$ of the displacement 
in force direction as a function of $\Delta t$
and require that analytical results for 
ABPs also hold for the higher moments if we use $\Delta \vec{r}_\text{EC}$
for the displacement and $\langle \Delta t_\text{EC} \rangle $ for $\Delta t$.

Simple exact analytical results for 
these higher moments for single free ABPs 
are not available, but approximate results which hold in the short time 
limit  $\Delta t \ll \tau$:
\begin{align}
	\langle (\Delta \vec{r} \cdot \vec{e})^2 \rangle &\approx (v_0 \Delta t)^2 +  2D\Delta t 
	\label{eq:eom1}, \\
	\langle (\Delta \vec{r} \cdot \vec{e})^3 \rangle &\approx (v_0 \Delta t)^3 +  6 v_0 D\Delta t^2
	\label{eq:eom2}, \\
	\langle \Delta \vec{r}^2 \rangle &\approx (v_0 \Delta t)^2 +  4D\Delta t \label{eq:eom3}.
\end{align}

We have calculated the algorithmic 
counterpart of the left hand sides of Eqs.\ \ref{eq:eom1}-\ref{eq:eom3} in 
Sec.\ \ref{sec: higher_moments} for one ECMC move analytically.
Figure \ref{fig:validation_kEC_algorithm_eom} shows $\langle (\Delta \vec{r}_{\mathrm{EC}} \cdot 
\vec{e})^2 \rangle/\ell_{\mathrm{EC}}^2$, $\langle (\Delta \vec{r}_{\mathrm{EC}} \cdot \vec{e})^3 \rangle/\ell_{\mathrm{EC}}^3$ and $\langle 
\Delta \vec{r}_{\mathrm{EC}}^2 \rangle/\ell_{\mathrm{EC}}^2$ as a function of the dimensionless parameter $x=\beta F_0
\ell_{\mathrm{EC}}$ and demonstrates perfect agreement between these analytical calculations (orange lines) 
and kEC simulation data (black crosses). Note that all analytical results of Sec.\ \ref{sec: higher_moments}
only depend on the parameter $x$.

It is not obvious whether or not Eqs.\ \ref{eq:eom1}-\ref{eq:eom3} for higher moments
are valid within the kEC algorithm  using the corresponding 
mean time per EC step~$\Delta t = \langle \Delta t_{\mathrm{EC}} \rangle$. 
For this reason Fig.\ \ref{fig:validation_kEC_algorithm_eom} also shows 
the right hand sides of 
Eqs.\ \ref{eq:eom1}-\ref{eq:eom3} for 
$\Delta t = \langle \Delta t_{\mathrm{EC}} \rangle$ (blue lines). 
We expect the approximations in Eqs.\ \ref{eq:eom1}\,-\,\ref{eq:eom3} to
break down for $\Delta t = \langle \Delta t_{\mathrm{EC}} \rangle > \tau$.
This is equivalent to 
$x> (3\pi /2) \mathrm{Pe}^2$ for high activity $\mathrm{Pe}>1$ 
(and $x\ll 2\sqrt{3} \mathrm{Pe}$   in the passive limit  $\mathrm{Pe}\ll 1$), see main text.  
 Numerically, we find
the values of $\Delta t$, at which the approximations in Eqs.\ \ref{eq:eom1}-\ref{eq:eom3} 
break down (= relative error of at least $5\%$ to the exact result), correspond - for all relevant 
active forces - to values of $x>x_c = 10^1 - 10^4$. In other words: the origin of the deviations in 
Fig.\ \ref{fig:validation_kEC_algorithm_eom} between the blue and orange lines 
are \emph{not} a result of the approximations in 
Eqs.\ \ref{eq:eom1}-\ref{eq:eom3}  itself but are solely algorithmic for $x < x_c$, i.e., due to a failure 
of  our choice $\Delta t = \langle \Delta t_{\mathrm{EC}} \rangle$ for a 
timestep.

Figure \ref{fig:validation_kEC_algorithm_eom}  shows that 
these algorithmic deviations between left hand and right hand sides of Eqs.\ 
\ref{eq:eom1}-\ref{eq:eom3} (orange and blue lines, respectively) 
become large for $x=\beta F_0 \ell_{\mathrm{EC}}>1-10$. 
For an optimal EC lengths $\ell_\text{EC}$, we would require left hand and right hand sides to 
exhibit a common intersection at a corresponding optimal $x$-value.
Since there are no common intersections, 
we are unable to determine a \emph{perfect} EC length, which would reproduce 
all quantities related to the equations of motion correctly on a
single particle level. However, in accordance with the observations made in the main text, 
all deviations in Fig.\ \ref{fig:validation_kEC_algorithm_eom} vanish 
for \emph{small} $x=\beta F_0 \ell_{\mathrm{EC}} \ll 10^{-1}$. This implies that all 
of those one-particle single-step quantities are correctly reproduced
in the limit~$\ell_{\mathrm{EC}} / \sigma  = x/3\mathrm{Pe}\rightarrow 0$. 
More generally, we will derive that, for a single ABP,   the kEC algorithm  
converges to  the correct one-particle Fokker-Planck equation  
in the limit of vanishing EC lengths in the following section.

\subsection{The one-particle Fokker-Planck equation for the kEC algorithm converges to the 
   ABP Fokker-Planck equation for small $\ell_{\mathrm{EC}}$}
\label{sec:derivation_one_body_fp_equation}

The stochastic motion of a single free ABP can be completely captured by the 
one-particle Fokker-Planck equation for the 
probability density  $P(\vec{r},\theta, t)$  to find the particle at position $\vec{r}$ with orientation 
$\theta$ at time $t$,
\begin{align}
		\partial_t P(\vec{r},\theta, t) 
		&\approx - (v_0 \vec{e}) \cdot \vec{\nabla}_{\vec{r}} P(\vec{r},\theta, t) 
		+ D_r \partial^2_{\theta} P(\vec{r}, \theta, t) \nonumber \\
		&~~ + D \vec{\nabla}_{\vec{r}}^2 P(\vec{r},\theta, t),	
		\label{eq:FPABP}
\end{align}
In the following we will show that the kEC algorithm converges to  the same 
one-particle Fokker-Planck equation  in the limit of vanishing EC lengths or 
small  timesteps $\Delta t = \langle \Delta t_{\mathrm{EC}} \rangle$. 
As a result, \emph{all} moments
$\langle (\Delta \vec{r} \cdot \vec{e})^n \rangle$ of the displacement as a function of 
$\Delta t$ are correctly reproduced in this limit further generalizing the results of the previous sections.

Let $P(\vec{r},\theta, t)$ be the probability density to find the particle at $(\vec{r}, \theta)$ at time 
$t$ and $p_\phi(\vec{r}|\vec{r}',\theta|\theta')$ the probability 
density to transition from $\vec{r}'$ to $\vec{r}$ and from  $\theta'$ to $\theta$ within one step of the kEC algorithm
with EC direction $\vec{e}_\text{EC}$. 
The EC direction is  given by the angle $\phi$ with  the unit propulsion 
direction vector $ \vec{e}$.
We average $P(\vec{r},\theta, t)$ over all angles $\phi$, i.e., over all EC directions in each kEC step until 
time $t$. 
The  probability density $P(\vec{r}, \theta, t + \langle \Delta t_\text{EC} \rangle)$ 
one kEC timestep later can be related to $P(\vec{r},\theta, t)$ via
\begin{align}
   & P(\vec{r}, \theta, t + \langle \Delta t_\text{EC} \rangle) \nonumber\\
     &= \frac{1}{2\pi} \int_{0}^{2\pi} d\phi \int_V d^2\vec{r}' \int_{-\infty}^\infty d\theta'  p_\phi(\vec{r}|\vec{r}',\theta|\theta')P(\vec{r}', \theta',t),
\end{align}
where we average over all angles $\phi$ also in the last kEC step. 
The average time corresponding to one 
kEC step  $\langle \Delta t_\text{EC} \rangle$; we will see in the course of our derivation of the 
Fokker-Planck equation how  the algorithmic timestep 
$\langle \Delta t_\text{EC} \rangle$ has to be chosen to arrive at  the 
ABP Fokker-Planck equation (\ref{eq:FPABP}). Analogously to the standard derivation of Fokker-Planck equations, 
we can argue that 
the transition probability $p_\phi(\vec{r}|\vec{r}',\theta|\theta')$ for a single kEC timestep is peaked 
around $\theta=\theta'$ and, typically,  $|\vec{r}-\vec{r}'| \sim \ell_\mathrm{EC}$
for one kEC step. We will consider the limit of small $\ell_\mathrm{EC}$ and 
will neglect variations on scales $|\vec{r}-\vec{r}'| \lesssim \ell_\mathrm{EC}$. 
Therefore, we can expand $P(\vec{r}', \theta',t)$ in $\vec{r}'$ and $\theta'$ around $\vec{r}$
and $\theta$ to obtain 
\begin{align}
   & P(\vec{r}, \theta, t + \langle \Delta t_\text{EC} \rangle) \nonumber\\
     &\approx  P(\vec{r}, \theta, t) + \frac{1}{2\pi} \int_{0}^{2\pi} d\phi \int_V d^2\vec{r}' \int_{-\infty}^\infty d\theta'  p_\phi(\vec{r}|\vec{r}',\theta|\theta') \times \nonumber\\
     &~\Big(  
     -\vec{\nabla}_\vec{r}P(\vec{r}, \theta,t) \cdot (\vec{r}-\vec{r}') \nonumber\\
     &~+\frac{1}{2} \sum_{i,j=1}^2\partial_{r_i}\partial_{r_j} P(\vec{r}, \theta,t)(r_i-r_i')(r_j-r_j')
     \nonumber\\
     &~ - \partial_\theta P(\vec{r}, \theta,t)(\theta-\theta')+ \frac{1}{2} \partial_\theta^2  P(\vec{r}, \theta,t)(\theta-\theta')^2 
     \Big)
     \nonumber\\
    &= P(\vec{r}, \theta, t) -  \vec{\nabla}_\vec{r}P(\vec{r}, \theta,t) \cdot \langle \Delta \vec{r}_{\mathrm{EC}}\rangle 
       \nonumber\\
    & ~+ \frac{1}{2}\partial_x^2 P(\vec{r}, \theta,t) \langle \Delta x^2_{\mathrm{EC}} \rangle
     + \frac{1}{2}\partial_y^2 P(\vec{r}, \theta,t) \langle \Delta y^2_{\mathrm{EC}} \rangle
        \nonumber\\
    & ~-  \partial_\theta P(\vec{r}, \theta,t)\langle \Delta \theta_{\mathrm{EC}}\rangle
      + \frac{1}{2} \partial_\theta^2  P(\vec{r}, \theta,t)\langle \Delta \theta_{\mathrm{EC}}^2\rangle.
       \label{eq:FPTaylor}
\end{align}

For our kEC algorithm the transition probability factorizes, $p_\phi(\vec{r}|\vec{r}',\theta|\theta') = 
f_\phi(\vec{r}|\vec{r}') g(\theta|\theta')$ because, 
after each  ECMC move,  the angular change $\Delta \theta$ is drawn  from a Gaussian distribution with zero mean 
$\langle \Delta \theta_{\mathrm{EC}}\rangle=0$ and 
and variance $\langle \Delta \theta_{\mathrm{EC}}^2\rangle = 2D_r \langle \Delta t_\text{EC} \rangle$ independently 
of the ECMC move vector. Therefore, there are no mixed terms in the Taylor expansion (\ref{eq:FPTaylor}).
We can specify the spatial transition probability as 
\begin{equation}
    f_\phi(\vec{r}|\vec{r}')\vert_{\cos\phi \geq 0} 
    = \delta\big(\vec{r} - (\vec{r}' + \ell_\text{EC} \vec{e}_\text{EC})\big)
\end{equation}
for $\cos\phi \geq 0$ and
\begin{align}
&    f_\phi(\vec{r}|\vec{r}')\vert_{\cos\phi < 0}
= \int_0^{\ell_{\text{EC}}} \text{d}x p_d(x) 
   \times \nonumber\\
   &~~~~~~~~\delta\big(\vec{r} - (\vec{r}'  
    + \ell_{\text{EC}} \vec{e}_{\text{EC}} 
    - 2 \cos \phi (\ell_{\text{EC}} - x) \vec{e})\big) 
    \nonumber\\
    &~+ \delta\big(\vec{r}-(\vec{r}'+\ell_{\text{EC}} \vec{e}_{\text{EC}})\big) e^{-\beta F_a 
    \ell_{\text{EC}} \vert \cos \phi \vert}
\end{align}
for $\cos\phi < 0$ with $p_d(x)$ previously defined in Eq.\ (\ref{eq:rejection_distance_distribution}).
Therefore, the transition probability $p_\phi(\vec{r}|\vec{r}',\theta|\theta')$ for a single kEC timestep is Gaussian in the angle $\theta$ and peaked 
around $\theta=\theta'$ but, obviously,  not Gaussian in $\vec{r}$ 
due to the deterministic nature of EC moves. Nevertheless, also the 
above results for $f_\phi(\vec{r}|\vec{r}')$ are peaked around 
$|\vec{r}-\vec{r}'| \sim \ell_\mathrm{EC}$
for one kEC step as assumed in  the Taylor expansion in 
Eq.\ (\ref{eq:FPTaylor}).

The Taylor expansion in $\theta$ is justified if 
$\langle \Delta \theta_{\mathrm{EC}}^2\rangle \ll 1$ or 
$\langle \Delta t_\text{EC} \rangle \ll 1/D_r = \tau$. This is exactly the condition of smooth rotation discussed 
in the main text. If this condition is fulfilled we can
also neglect higher orders of the Taylor expansion in $\theta$ because 
$\langle \Delta \theta_{\mathrm{EC}}^{2n}\rangle\sim (D_r \langle \Delta t_\text{EC} \rangle)^n$.
Typically this is a relatively weak condition requiring only   moderately small EC lengths
$\ell_\text{EC}/\sigma \ll (\pi /2) \mathrm{Pe}$ for high activity $\mathrm{Pe}>1$.
In Eq.\ (\ref{eq:FPTaylor}),  we also neglected higher orders of the Taylor expansion 
in $\vec{r}$ because
$\langle \Delta r^n_{i,\mathrm{EC}} \rangle\sim \ell_{\mathrm{EC}}^n$
will be justified by the requirement of small EC lengths.

Further approximating $\partial_t P(\vec{r}, \theta, t)\approx 
(P(\vec{r}, \theta, t + \langle \Delta t_\text{EC} \rangle)-P(\vec{r}, \theta, t))/\langle \Delta t_\text{EC} \rangle$
and using
\begin{align}
   \langle \Delta x^2_{\mathrm{EC}} \rangle =   \langle \Delta y^2_{\mathrm{EC}} \rangle 
   &\approx \frac{1}{2} \ell_{\mathrm{EC}}^2   ~~(x\ll 1).
\end{align}
for the second moments of the ECMC move length (see Sec.\ \ref{sec: higher_moments}), 
we obtain 
\begin{align}
   \partial_t P(\vec{r}, \theta, t) \nonumber
     \approx  &- \vec{\nabla}_\vec{r}P(\vec{r}, \theta,t) \cdot 
     \frac{\langle \Delta \vec{r}_{\mathrm{EC}}\rangle }{\langle \Delta t_\text{EC} \rangle}
       \nonumber\\
    & ~+ 
       \vec{\nabla}_\vec{r}^2P(\vec{r}, \theta,t)  
    \frac{ \ell_{\mathrm{EC}}^2}{4\langle \Delta t_\text{EC} \rangle} 
     \nonumber\\
    & ~
      + D_r \partial_\theta^2  P(\vec{r}, \theta,t).
       \label{eq:FPTaylor2}
\end{align}
At this point, we see that the algorithmic timestep 
$\langle \Delta t_\text{EC} \rangle$ has to be chosen as 
\begin{align}
    \langle \Delta \vec{r}_{\mathrm{EC}}\rangle &= 
        \vec{v}_0  \langle \Delta t_\text{EC} \rangle
\end{align}
for the convective term  to agree with the 
ABP Fokker-Planck equation (\ref{eq:FPABP}).
This is exactly our  definition of the mean time $\langle \Delta t_\text{EC} \rangle$
from the main text. 
Finally, if also the diffusive term  is required to agree with the 
ABP Fokker-Planck equation (\ref{eq:FPABP}), we need
\begin{align}
    \frac{ \ell_{\mathrm{EC}}^2}{4\langle \Delta t_\text{EC} \rangle}   &\approx  D .
    \label{eq:strong}
\end{align}
This poses a relatively strong condition and is only fulfilled for $x\ll 1$ or small EC lengths
$\ell_\text{EC}/\sigma \ll 1/3\mathrm{Pe}$. For $x>1$, the apparent translational 
diffusion constant $D'={ \ell_{\mathrm{EC}}^2}/{4\langle \Delta t_\text{EC} \rangle} =D/4f(x)$ is  \emph{larger}
than $D$ 
because $f(x)$ is a decreasing function of $x$. 
If the condition (\ref{eq:strong}) is fulfilled, the one-particle  Fokker-Planck equation 
of the kEC algorithm agrees with the ABP Fokker-Planck equation, assuring the validity of the kEC algorithm 
for single particles. 

The strong condition $x\ll 1$ can be weakened in practice because the diffusive term of the Fokker-Planck equation 
is only relevant for small time scales.  There are three dynamic regimes for ABPs: (i) on time scales $\Delta t< D/v_0^2$ we find translation diffusion  with $\langle \Delta r^2 (\Delta t)\rangle \sim D\Delta t$; (ii) on intermediate time scales $D/v_0^2<\Delta t< \tau $ below the persistence time, the ABP performs ballistic motion with
$\langle \Delta r^2 (\Delta t)\rangle \sim v_0^2\Delta t^2$;
(iii) on large time scales $\Delta t> \tau$, 
the motion becomes a persistent random walk with 
$\langle \Delta r^2 (\Delta t)\rangle \sim v_0^2\tau \Delta t$.
As a consequence, a violation of condition (\ref{eq:strong}) will only give rise to a wrong translational diffusion constant that affects the short-time translational diffusion regime (i).
The apparent diffusion constant 
$D'=D/4f(x)$ is  larger  than $D$ and resulting in  
wrong results on time scales below the corresponding crossover time scale $\Delta t < D'/v_0^2$.
As long as this erroneous  behavior is limited to length scales $\langle \Delta r^2 (\Delta t)\rangle < \sigma^2$, it will affect only small scale features of the distribution  $P(\vec{r}, \theta, t)$ corresponding to variations below one particle diameter $\sigma$ (or  Fourier modes of the distribution with wave vectors $k\gtrsim 1/\sigma$). 
This corresponds to small time scales $\Delta t < \sigma/v_0$, and  as long as  $\sigma/v_0> D'/v_0^2$, only
features of the distribution below the small scale $\sigma$ 
are wrong. This condition can be written as 
$3{\rm Pe} = v_0\sigma /D > D'/D =1/4 f(x) $ resulting in the  weaker condition 
\begin{equation}
 x<24{\rm Pe}/\pi  ~~\mbox{or}~~  \ell_\text{EC}/\sigma < 8/\pi.
 \label{eq:weak}
\end{equation}

In the derivation of the Fokker-Planck equation we neglected  variations  of the distribution  $P(\vec{r}, \theta, t)$ on scales 
$\Delta r \lesssim \ell_\mathrm{EC}$, which remains consistent with neglecting variations 
on scales $\Delta r \lesssim \sigma$ if the condition  $\ell_\text{EC}/\sigma < 8/\pi$ holds.
Under this condition,  we can also neglect higher orders of the Taylor expansion 
in $\vec{r}$ in Eq.\ (\ref{eq:FPTaylor}) because
$\langle \Delta r^n{i,\mathrm{EC}} \rangle\sim \ell_{\mathrm{EC}}^n \lesssim \sigma^n$, 
which is also consistent with neglecting variations 
on scales $\Delta r < \sigma$. 
We conclude that we can increase $\ell_\text{EC}/\sigma$ up to order unity 
if we are not interested in features of the distribution  $P(\vec{r}, \theta, t)$ 
below one particle diameter $\sigma$.

\subsection{Two-body problem}

\begin{figure}
	\includegraphics[width=1\linewidth]{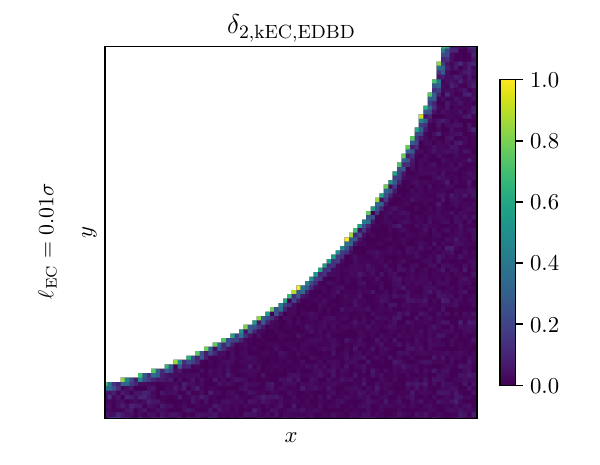}
	\caption{
		A cutout section from the heatmap of the absolute value of the 
		relative error per grid element 
		$\delta_{\text{2,kEC,EDBD}}$ 
		of the 	two-particle  probability density $p_2(\vec{r},t)$ 
		between kEC and EDBD algorithms
		using $\ell_{\mathrm{EC}} = 0.01\sigma$ for the second 
		initial configuration in Fig.\ \ref{fig:two_body_problem} ($t=0.05\tau$). 
		Even though the deviations are in general relatively small, it is notable that the
		error per grid element seems to increase if both particles are close to each other.
		}
	\label{fig:additional_plot_two_body_problem_validation}
\end{figure}

Regarding the two-body problem, Fig.\ \ref{fig:additional_plot_two_body_problem_validation} 
presents a section of the heatmap of the absolute value of the relative deviation per grid element 
$\delta_{\text{2,kEC,EDBD}}$
of the 	two-particle  probability density $p_2(\vec{r}_b,t)$ 
between kEC and EDBD algorithms for 
$\ell_{\mathrm{EC}} = 0.01\sigma$ and $\tau_B = 0.01$ from the second 
configuration in Fig.\ \ref{fig:two_body_problem} ($t=0.05\tau$). 
One may notice the relative errors of the grid elements in close proximity to 
the coordinate-fixed particle are seemingly higher than the ones further away.
Assuming the EDBD algorithm  to be correct, this may suggest that 
the kEC algorithm is not  able to  handle 
interactions between two particles completely correctly.

This objection, or at least its relevance, can be dismissed: 
Similar to the role of the displacement
length for the kEC algorithm, the EDBD algorithm depends on a parameter $\tau_B$ 
(for further information on this subject see Ref.\ \cite{Levis2017}), 
whose choice influences the accuracy of the algorithm and is only exact in the limit 
of~$\tau_B \rightarrow 0$. As mentioned before, we cannot prove our algorithm 
to be exact, but presumably decreasing both $\ell_{\mathrm{EC}}$ 
and $\tau_B$ even further will lead to the deviations in proximity to the 
coordinate-fixed particle to shrink. If this is true, we can (in theory) tune 
both the deviations themselves as well as the length scales, on which they 
are relevant, at will. In practice this is of course limited to a 
certain degree since the required wall time 
increases, if the displacement length is reduced.

\section{Details on the time distribution in many-body simulations}
\label{sec:time_distribution}

\begin{figure}[ht]
    \includegraphics[width=0.4\linewidth]{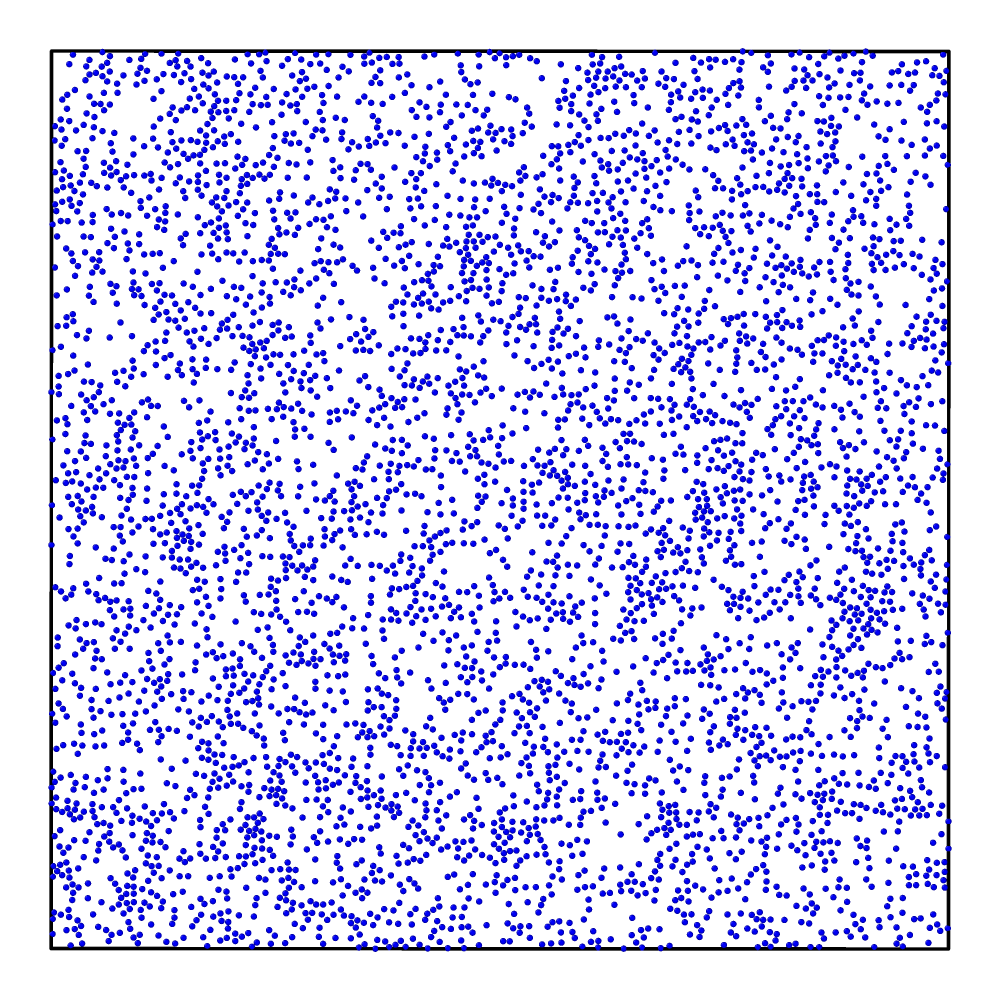}
    \includegraphics[width=0.4\linewidth]{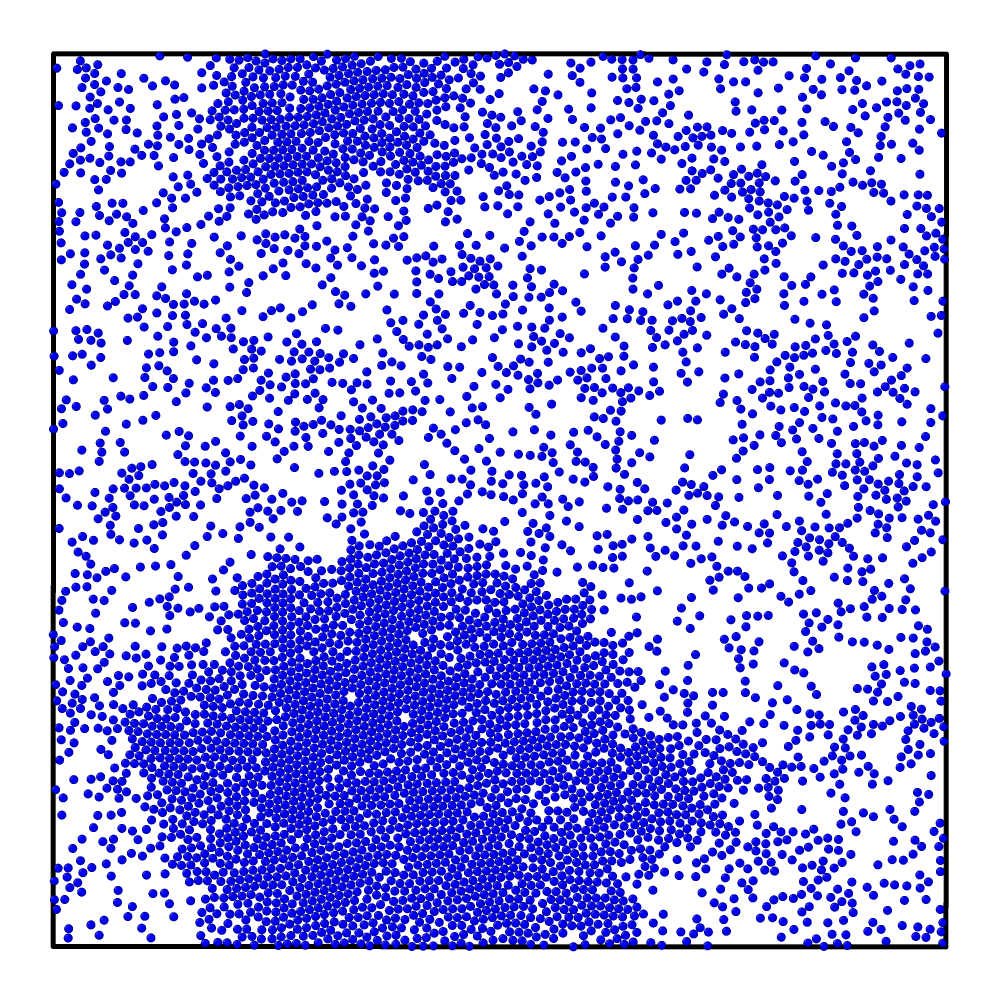}
	\caption{
        Exemplary snapshots of the two systems, which were analyzed to gain insight about 
	    the behaviour of the time distribution. The system on the left with $\mathrm{Pe} 
		= 10, \eta = 0.20, N=5000$ is dilute and MIPS does not occcur. The system on the 
    	right with  $\mathrm{Pe} = 20, \eta = 0.40, N=5000$ is dense and exhibits MIPS. 
	    We note that system sizes are different for both system (but shown scaled to the 
		same size), accordingly the spheres in the dense system on the right appear larger. 
	}
	\label{fig:snapshots_time_distribution}
\end{figure}

\begin{figure}[ht]
    \includegraphics[width=\linewidth]{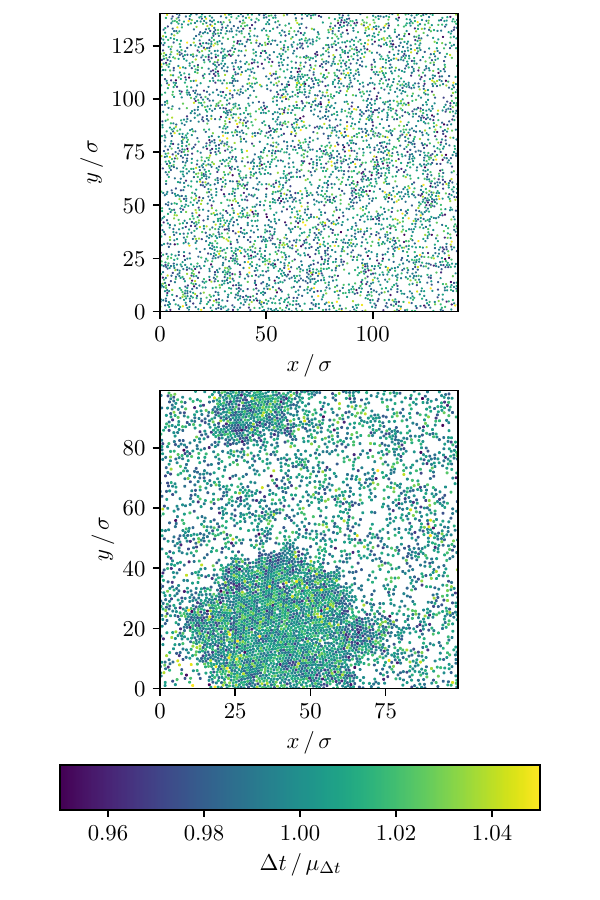}
    \caption{
        Snapshot of the two systems mentioned in Fig.\ \ref{fig:snapshots_time_distribution} 
        (upper: $\mathrm{Pe} = 10, \eta = 0.20, N=5000$, 
        lower: $\mathrm{Pe} = 20, \eta = 0.40,
        N=5000$) color-coded with respect to the individual particle times $\Delta t$ 
        at $\mu_{\Delta t} = 100\tau$ using $\ell_{\mathrm{EC}} = 1.00\sigma$. 
        Regardless of the presence of MIPS 
        no spatial difference in the individual particle times can be observed, 
        implying the
        same time passes in each phase.
	}
	\label{fig:color_coded_snapshots_time_distribution}
\end{figure}

\begin{figure}
	\includegraphics[width=1\linewidth]{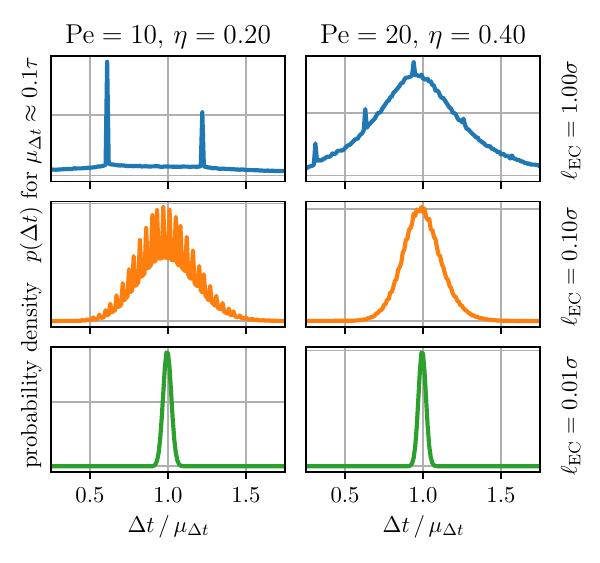}
	\caption{
		Time distribution $p(\Delta t)$ depending on the EC length for  the dilute system without MIPS 
		on the left and the dense system that exhibits MIPS on the right. We focus on short timescales 
		as the mean time passed per particle is~$\mu_{\Delta t} = 0.1\tau$. For decreasing
		$\ell_{\mathrm{EC}}$ the time distribution function $p(\Delta t)$ becomes a sharp bell-like 
		function. Therefore the kEC algorithm can be set up in a way that for every particle almost the 
		same time has passed.
		}
	\label{fig:time_distribution_short_time_scales}
\end{figure}

\begin{figure}
	\includegraphics[width=1\linewidth]{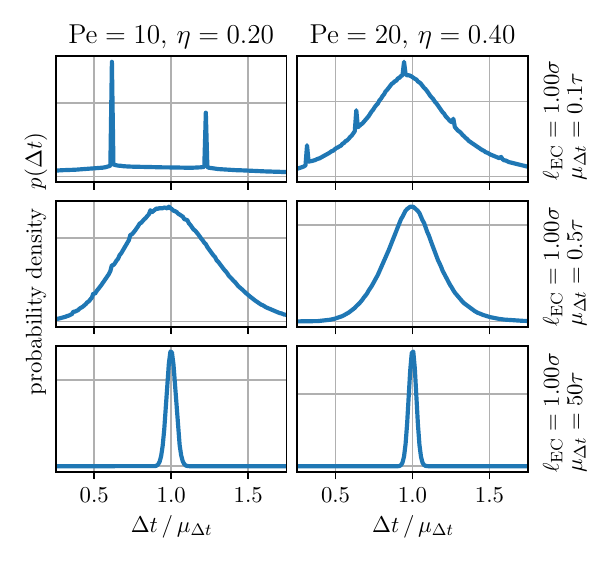}
    \caption{
		Time distribution $p(\Delta t)$ for the dilute system without MIPS 
		on the left and the dense system that exhibits MIPS on the right.  
	    We focus on medium (mean time passed per particles is~$\mu_{\Delta t} = 0.5\tau$) and long ($\mu_{\Delta t} = 50\tau$) timescales.
		For increasing $\mu_{\Delta t}$ the time distribution function 
		$p(\Delta t)$ becomes a bell-like function, whose relative standard deviation
		$c_{\Delta t} = \sigma_{\Delta t} / \mu_{\Delta t}$ steadily decreases. 
		The apparent 
		convergence of the relative values of the individual particle times 
		in the limit $\mu_{\Delta t} \rightarrow \infty$ is key to 
		satisfy the requirement that for all particles the same time must have passed.
		}
	\label{fig:time_distribution_long_time_scales}
\end{figure}

\begin{figure}
	\includegraphics[width=1\linewidth]{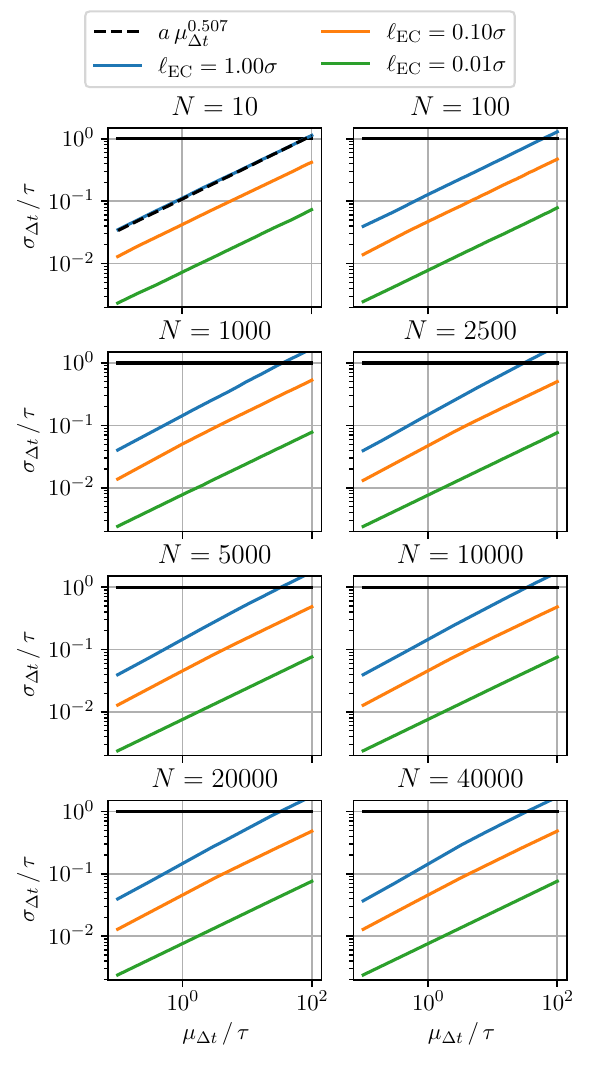}
    \caption{
        Relationship between the mean time per particle and
        the standard deviation of the time distribution for
        different system sizes and displacement lengths
        $\ell_{\mathrm{EC}}$ ($\mathrm{Pe} = 20$ and
        $\eta = 0.40$).
	}
	\label{fig:mean_time_and_standard_deviation_kEC}
\end{figure}

As previously stated every particle has its \textit{own individual} 
simulation time $t$. This may cause problems - 
especially in a many-body simulation - because 
we would naively require that for all particles the exact  
same time has passed in order to maintain the correct propulsion 
directions (which are undergoing rotational diffusion with a persistence time 
$\tau$), which is a necessary condition to also preserve the order and 
causality of collisions (or interactions in general).
However, in statistical physics, where we are interested in 
statistical averages and, typically, also averages over all particles 
(for densities of extensive observables), this is not necessary, 
and it appears to be sufficient if fluctuations in time do not cause uncontrollable 
statistical fluctuations of observables.
There are three types of averages that we have to consider in any MC or molecular dynamics simulation:
(1) Statistical  averages $\langle O(\Delta t) \rangle$ 
of a  quantity  $O(\Delta t)$ depending on a time difference $\Delta t$ 
can be obtained 
by taking averages over many samples of the system, each after 
the specified time difference $\Delta t$.
(2) Averages $\langle O \rangle$ 
of a time-independent quantity  $O$ can also be 
calculated as time averages over long time scales. 
(3) If $O = N^{-1} \sum_i O_i$ is the density of an extensive quantity,
we can average over all particles, in addition to sample averaging (1) or
time averaging (2).
We have to check for all three types of averages how they are affected by 
particles having slightly different, \textit{individual} times in the 
kEC simulation technique. 
In particular, the  approach (2) to take time averages 
has to be applied with care.

To control the errors of any sample, time, or particle average, we have to 
investigate the time distribution of particles for different EC lengths, 
timescales, system sizes, and different phases of the system such as the 
dense MIPS phase or a dilute phase. We focus on two  systems of $N=5000$ 
particles in different parameter regimes: (i) a dilute 
system~($\mathrm{Pe} = 10, \eta = 0.20$), where MIPS does not occur,
and (ii) a dense system ($\mathrm{Pe} = 20, \eta = 0.40$), where MIPS 
does occur. Exemplary snapshots can be found in 
Fig.\ \ref{fig:snapshots_time_distribution}.

First of all, we want to investigate the distribution of passed times $\Delta t$ 
of the particles in
the dense system snapshot just mentioned. In Fig.\ \ref{fig:color_coded_snapshots_time_distribution} 
one can find the particles color-coded with respect to their individual time. It is 
worth noting in the case MIPS occurs one cannot find a systematic difference between 
the individual times of the particles in the dilute phase and the dense one. This is 
significant, as it proves the same time passes in the entire system regardless of the 
local density.

In Fig.\ \ref{fig:time_distribution_short_time_scales}, we show the distributions 
$p(\Delta t)$ of particle times $\Delta t$ on short timescales~(mean time per 
particle $\mu_{\Delta t} = 0.1\tau$): regardless of whether MIPS occurs the 
time distributions that are measured using large EC lengths are rather wide 
and everything but sharp. Additionally one can observe multiple equidistant 
maxima, separated by the mean time per EC step $\langle \Delta t_{\mathrm{EC}} 
\rangle$. These maxima appear particularly in the dilute phase, if no collisions 
and subsequent lifting moves take place during one EC move, which results in 
the total time $\langle \Delta t_{\mathrm{EC}} \rangle$ being assigned to each 
single particle move according to Eq.~(\ref{eq:tEC}).
Reducing the EC length leads to an increase in the number of maxima. The reason for 
this is related to the non-linear relationship between $\langle \Delta t_{\mathrm{EC}} 
\rangle$ and the EC length. By decreasing $\ell_{\mathrm{EC}}$ the mean time per EC 
step also gets smaller and therefore the number of EC steps per time interval increases. 
This implies that each particle is at least on average more often the starting point
of a new EC step and due to the increase of the total displacement length more lifting
moves take place. Consequently both the relative and absolute height of each peak shrinks.
Because collisions (and therefore lifting moves) are in general way more frequent in a 
dense system, smooth shapes of the time distribution are reached for larger EC lengths 
as it is less probable to assign $\langle \Delta t_{\mathrm{EC}} \rangle$ to a single 
particle. Finally, reducing the EC length even further leads to bell-shaped time distributions, 
centered around the mean time per particle~$\mu_{\Delta t}$, whose standard 
deviation~$\sigma_{\Delta t}$ appears to follow a power law $\sigma_{\Delta t}\propto \ell_{\mathrm{EC}}^{\alpha}$ with an exponent $\alpha>0$. The shape of these distributions
is in some way to be expected, as we anticipate the total displacement length of each particle 
and therefore also the individual time to be approximately the same if the total number of EC 
steps and therefore the total displacement length is large.

Concerning medium~($\mu_{\Delta t} = 0.5\tau$) and 
long~($\mu_{\Delta t} = 50\tau$) timescales 
(see Fig.\ \ref{fig:time_distribution_long_time_scales}), we notice - 
regardless of the particle density - the relative and absolute 
values of the equidistant peaks separated by the mean time per EC 
step $\langle \Delta t_{\mathrm{EC}} \rangle$ 
to decrease steadily by increasing 
the mean time per particle $\mu_{\Delta t}$. 
This appears to be a rather fast process, 
as the maxima are barely visible already 
at $\mu_{\Delta t} = 0.5\tau$. Additionally and 
similar to reducing the EC length $\ell_{\mathrm{EC}}$, 
a growing mean time per 
particle leads to the time distribution to become 
bell-shaped. Furthermore the relative standard deviation 
$\sigma_{\Delta t} / \mu_{\Delta t}$ that corresponds to the visible width in 
the plots in Fig.\ \ref{fig:time_distribution_long_time_scales}, decreases 
as a power law $\sigma_{\Delta t}/ \mu_{\Delta t} \propto \mu_{\Delta t}^{-1/2}$ 
as $\mu_{\Delta t}$ increases, 
implying the individual times of the particles are converging relatively.

Based on these observations, we investigate the relationship 
between the mean time per particle and 
the standard deviation of the time distribution and the 
influence of both the number of particles $N$ 
as well as the EC length for the dense system 
$\mathrm{Pe} = 20$ and $\eta = 0.40$. 
Fig.\ \ref{fig:mean_time_and_standard_deviation_kEC} reveals the standard 
deviation $\sigma_{\Delta t}$ of the time distribution to be 
proportional to the square root of the 
mean time per particle, $\sigma_{\Delta t} \propto \mu_{\Delta t}^{1/2}$. 
Additionally one may notice $\sigma_{\Delta t}$ to initially 
increase with $N$, reach an EC length dependent peak 
($N_{\mathrm{peak}} = 2500$ 
for $\ell_{\mathrm{EC}} = 1.00\sigma$, $N_{\mathrm{peak}} = 1000$ 
for $\ell_{\mathrm{EC}} = 0.10\sigma$ and $N_{\mathrm{peak}} = 100$ 
for $\ell_{\mathrm{EC}} = 0.01\sigma$) and afterwards to drop slightly, 
settle and thus become independent of $N$.
We find 
\begin{equation}
    \sigma_{\Delta t} = a(N)\ell_\text{EC}^\alpha \mu_{\Delta t}^{1/2}
    \label{eq:sigmat}
\end{equation}
with $a$ only weakly depending on $N$ and the exponent of $\ell_{\mathrm{EC}}$ ranging 
from $\alpha \approx 0.46$ for $N = 10$ to $\alpha \approx 0.56$ for $N = 40000$.

As mentioned before we require the individual particle times to roughly 
coincide, which can be quantified by the loosely defined criterion 
$\sigma_{\Delta t} < \tau$, which assures that particles remain within 
one persistence time, i.e., the propulsion directions will not be affected by 
the time spread. For $\ell_{\mathrm{EC}} = 0.01\sigma$, 
the standard deviation satisfies $\sigma_{\Delta t} \approx 0.1\tau$ 
at $\mu_{\Delta t} = 100\tau$ and clearly fulfills this criterion. 
This implies further that we can 
simulate the system for several thousand
persistence times using the kEC algorithm with
$\ell_{\mathrm{EC}} =0.01\sigma$ without ever violating the criterion.

The result (\ref{eq:sigmat}) enables us to discuss the three types of statistical averages mentioned 
in the beginning of the section systematically: First, we notice that a width $\sigma_{\Delta t}$ of 
the time distribution also implies measurement errors of an observable $\Delta O(\Delta t) \sim 
\sigma_{\Delta t}$ after a time $\Delta t$. (1) Averages over $N_s$ samples of a time-dependent 
quantity $\langle O(\Delta t) \rangle= N_s^{-1} \sum_{s=1}^{N_s} O_s(\Delta t)$ are based on the 
fact that the mean over $N_s$ samples has an error $\sigma_{\langle O(\Delta t) \rangle} 
=  N_s^{-1/2}\Delta O(\Delta t) \sim  N_s^{-1/2}\sigma_{\Delta t}$, which can be controlled by the 
sample number as long as samples are statistically independent. The result of the additional time 
distribution of particles in the kEC simulation is simply that we have to average over more 
samples $N_s$ for larger $\Delta t$ to reduce the error of the mean. (2) Time averages over $N_t$ 
times  of a time-independent quantity $\langle O\rangle= N_t^{-1} \sum_{n=1}^{N_t} O(n \delta t)$ in 
intervals $\delta t$ that should be larger than one autocorrelation time of the system have to be 
treated with more care because errors $\sigma_{n\delta t} \propto n^{1/2}$ grow in time according 
to Eq.\ (\ref{eq:sigmat}). Therefore, the error $\sigma_{\langle O(\Delta t) \rangle} = (N_t^{-2}
\sum_{n=1}^{N_t} \sigma^2_{n\delta t} )^{1/2} \propto N_t^0$ of such a time average is no longer 
reducing $\propto N_t^{-1/2}$ as usual, but actually saturating to an $N_t$-independent value. 
(3) Averages $O(\Delta t) = N^{-1} \sum_i O_i(\Delta t)$ over all particles will still be effective 
and reducing the error of the average $\sigma_{O(\Delta t)}  \propto N^{-1/2}$ because $\Delta 
O(\Delta t) \sim \sigma_{\Delta t}$ is essentially $N$-independent according to Eq.\ (\ref{eq:sigmat}).

Taking all of these information into account, our algorithm appears to be suitable 
to deal with many-body simulations. One can simply decrease the displacement length 
$\ell_{\mathrm{EC}}$ to align the individual particle times, both their relative 
and also absolute values. This allow us to satisfy the condition $\sigma_{\Delta t} 
< \tau$ for very long simulation times. Typically, we only need simulation times 
of $20-50$ persistence times to reach stationary states (in case of $N = 5000$) and 
observe phenomena such as MIPS. Therefore we can set up the kEC algorithm in a way 
that for every particle (almost) the exact same time passes. This is true regardless 
of the presence of MIPS. 
Moreover, averaging over many samples or over particles are still effective ways to 
reduce errors of mean values. Only time averages have to be treated with care as errors 
of mean values saturate (with respect to the number of samples). Errors of all 
averages, including time averages, can be controlled by shrinking $\ell_\text{EC}$.
To avoid confusion, it should be emphasized that statistical averages - solely taking
the time distribution into account - can only converge to correct values in the 
limit $\ell_\text{EC} \rightarrow 0$. For large displacement lengths some of the mean 
values are converging, but to values related to distorted dynamics.

\section{Derivation of the $N$-body Fokker-Planck equation in the dilute limit}
\label{sec:nbodyfokkerplanck_dilute}

In order to prove our kEC algorithm also to be 
exact for general $N$-body problem there are two
problems, which one needs to overcome: (1) different individual times as a consequence of the time
distribution and (2) complex algorithmic dynamics due to lifting moves. While we are currently not
able to handle the second problem, one can ignore this issue in the dilute limit $\eta \rightarrow 0$.
Admittedly, the system does not exhibit interesting  physical phenomena such as MIPS  in a pure dilute state. Nevertheless, we still want to explore this limit, 
because one may argue, if our algorithm is 
correct for the dilute $N$-body system it does not become discontinuously incorrect, if the packing fraction
is increased to finite values.

Before dealing with the $N$-body Fokker-Planck equation itself, we need to make sure the exact same 
time passes for every particle. Let us say, we want our system to evolve for the time interval $\Delta t$
($\Delta t / \tau \ll 1$). For a given EC length, we would require $m$ EC steps for a single particle to
satisfy
\begin{equation}
    \Delta t = m \langle \Delta t_{\mathrm{EC}} \rangle
    \label{eq:steps_per_sweep}
\end{equation}
and therefore $N \cdot m$ steps so that for each particle the time evolves on average by $\Delta t$. After 
$N \cdot m$ steps, the time distribution of the entire system - as demonstrated in the previous section - 
would be a Gaussian. We expect the standard deviation of the time distribution to shrink if one reduces the 
EC length, similar to the configuration we discussed in Sec.\ \ref{sec:time_distribution}.
In other words: $\ell_{\mathrm{EC}} \rightarrow 0$ automatically leads to 
$\sigma_{\Delta t} \rightarrow 0$. However, a Gaussian distribution with vanishing standard deviation is a
$\delta$-distribution
\begin{equation}
    \lim_{\sigma_{\Delta t} \rightarrow 0} \mathrm{Gaussian}(\mu_{\Delta t}, \sigma_{\Delta t}) =
    \delta(t - \mu_{\Delta t}).
\end{equation}
This way, we can make sure that all individual particle times are exactly the
same.

Let 
$P(\vec{r}_1,...,\vec{r}_N, \theta_1,..., \theta_N, t_1,...,t_N)$ 
be the probability density of the 
dilute (non-interacting) $N$-body system. In the limit $\ell_{\mathrm{EC}} \rightarrow 0$ all individual 
times are the same and $t_i = t$ holds. To find the $N$-body Fokker-Planck equation we need to find an 
expression for 
$P(\{\vec{r}_i\}, \{\theta_i\}, t + \Delta t)\equiv P(\vec{r}_1,...,\vec{r}_N, \theta_1,...,\theta_N, t + \Delta t)$. 
In Sec.\ \ref{sec:derivation_one_body_fp_equation}
we found
\begin{align}
    &P(\vec{r},\theta, t + \langle \Delta t_{\mathrm{EC}} \rangle) \nonumber\\ 
    &~\approx \, P(\vec{r},\theta, t) 
    -v_0 \langle \Delta t_{\mathrm{EC}} \rangle \vec{e}(\theta)
    \vec{\nabla}_{\vec{r}} P(\vec{r},\theta, t)
    \nonumber\\ 
    &~+ D_r \langle 
    \Delta t_{\mathrm{EC}} \rangle 
    \partial_{\theta}^2 P(\vec{r},\theta, t) 
    + D \langle \Delta t_{\mathrm{EC}} \rangle
    \vec{\nabla}_{\vec{r}}^2 P(\vec{r},\theta, t).
\end{align}
We can use this formula iteratively and ignore all terms 
smaller than $\mathcal{O}(\langle \Delta t_{\mathrm{EC}}
\rangle)$, which leads to 
\begin{align}
    &P(\vec{r},\theta, t + m \langle \Delta t_{\mathrm{EC}} \rangle) \nonumber\\ 
    &~\approx \, P(\vec{r},\theta, t) 
    -v_0 m \langle \Delta t_{\mathrm{EC}} \rangle \vec{e}(\theta)
    \vec{\nabla}_{\vec{r}} P(\vec{r},\theta, t)
    \nonumber\\ 
    &~+ D_r m \langle 
    \Delta t_{\mathrm{EC}} \rangle 
    \partial_{\theta}^2 P(\vec{r},\theta, t) 
    + D m \langle \Delta t_{\mathrm{EC}} \rangle
    \vec{\nabla}_{\vec{r}}^2 P(\vec{r},\theta, t).
\end{align}
In the dilute (non-interacting) limit all coordinates are completely independent of each other. This allows us
to employ the same procedure, we just used for a single ABP, 
to obtain
\begin{align}
    &P(\{\vec{r}_i\}, \{\theta_i\}, t + \Delta t) 
    = P(\{\vec{r}_i\},\{\theta_i\}, t + m \langle t_{\mathrm{EC}} \rangle)
   \nonumber\\
&~    \approx P(\{\vec{r}_i\},\{\theta_i\}, t)  \nonumber\\
    &~+ m \langle t_{\mathrm{EC}} \rangle \sum_{i = 1}^N \Big( -v_{0,i} \vec{e}_i \vec{\nabla}_{\vec{r}_i} + D_r 
    \partial^2_{\theta_i}
    + D \vec{\nabla}^2_{\vec{r}_i} \Big)
    \nonumber\\
&~~~~    P(\{\vec{r}_i\},\{\theta_i\}, t).
\end{align}
This can easily be transformed to
\begin{align}
    &\partial_t P(\{\vec{r}_i\}, \{\theta_i\}, t)
    \nonumber\\
&~    \approx \sum_{i = 1}^N \Big( -v_{0,i} \vec{e}_i \vec{\nabla}_{\vec{r}_i} + D_r 
    \partial^2_{\theta_i} + D \vec{\nabla}^2_{\vec{r}_i} \Big)
    P(\{\vec{r}_i\},\{\theta_i\}, t),
\end{align}
which we identify as the correct $N$-body Fokker-Planck equation of non-interacting ABPs.

\section{Details of the kEC simulations for the phase diagram of hard disks}

To construct the large phase diagram in Fig.\ \ref{fig:full_mips_phase_diagram}
in the main text, we simulate $N = 19500$ particles using
$\ell_{\mathrm{EC}} = 1.00\sigma$ in a square box with area $A = L^2$ 
and periodic boundary conditions. We change the global
packing fraction $\eta = N\pi (\sigma/2)^2/A$ 
by changing the system size $L$ at a fixed particle number. 
Depending on the P\'{e}clet number
$\mathrm{Pe} = v_0 /(\sigma D_r) = L_p/\sigma$ and the global 
packing density, the phase behavior is classified and the binodal is determined.

\subsection{Calculating the distribution of the local packing fraction}
\label{sec:calculation_distribution_local_packing_fraction}

\begin{figure}
  \begin{center}
  \includegraphics[width=0.40\textwidth]{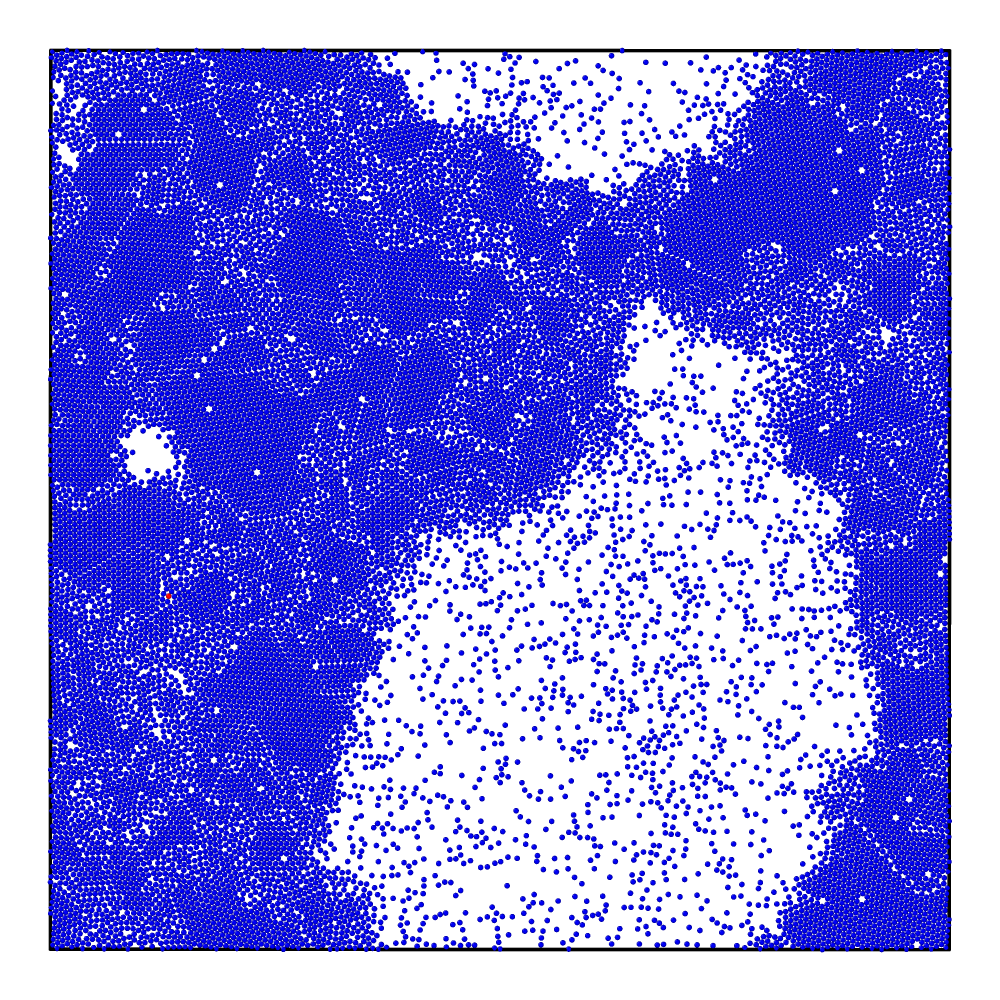}
  \includegraphics[width=0.47\textwidth]{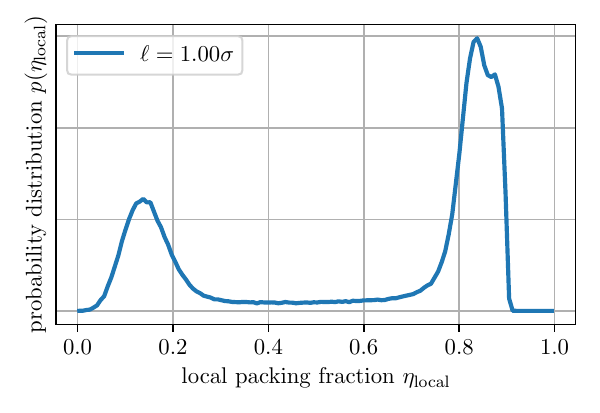}
  \caption{
   (a) Snapshot of a system exhibiting MIPS at $\mathrm{Pe} = 30$ and $\eta=0.60$ ($N=19500$, 
   	$\ell_\text{EC}=1.00\sigma$).
   (b) Corresponding histogram of local packing fractions $\eta_\text{local}$ using cells of 
   	area $A_h \approx (7\sigma)^2$.}
\label{fig:histogram}
\end{center}
\end{figure}

The investigation of the MIPS phase and its stability is performed via 
calculating the probability density distribution 
$p(\eta_{\mathrm{local}})$ of the
local packing fraction, see Fig.\ \ref{fig:histogram}.
In order to calculate this quantity, we first 
divide the simulation volume into multiple subvolumes, each of which has the 
area $A_h \approx (7\sigma)^2$.
We calculate the local density in each of those subvolumes and 
(by averaging over many snapshots) use
this data to construct the probability density distribution 
$p(\eta_{\mathrm{local}})$ of the
local packing fraction.

\begin{figure}
	\begin{center}
	\includegraphics[width=0.47\textwidth]{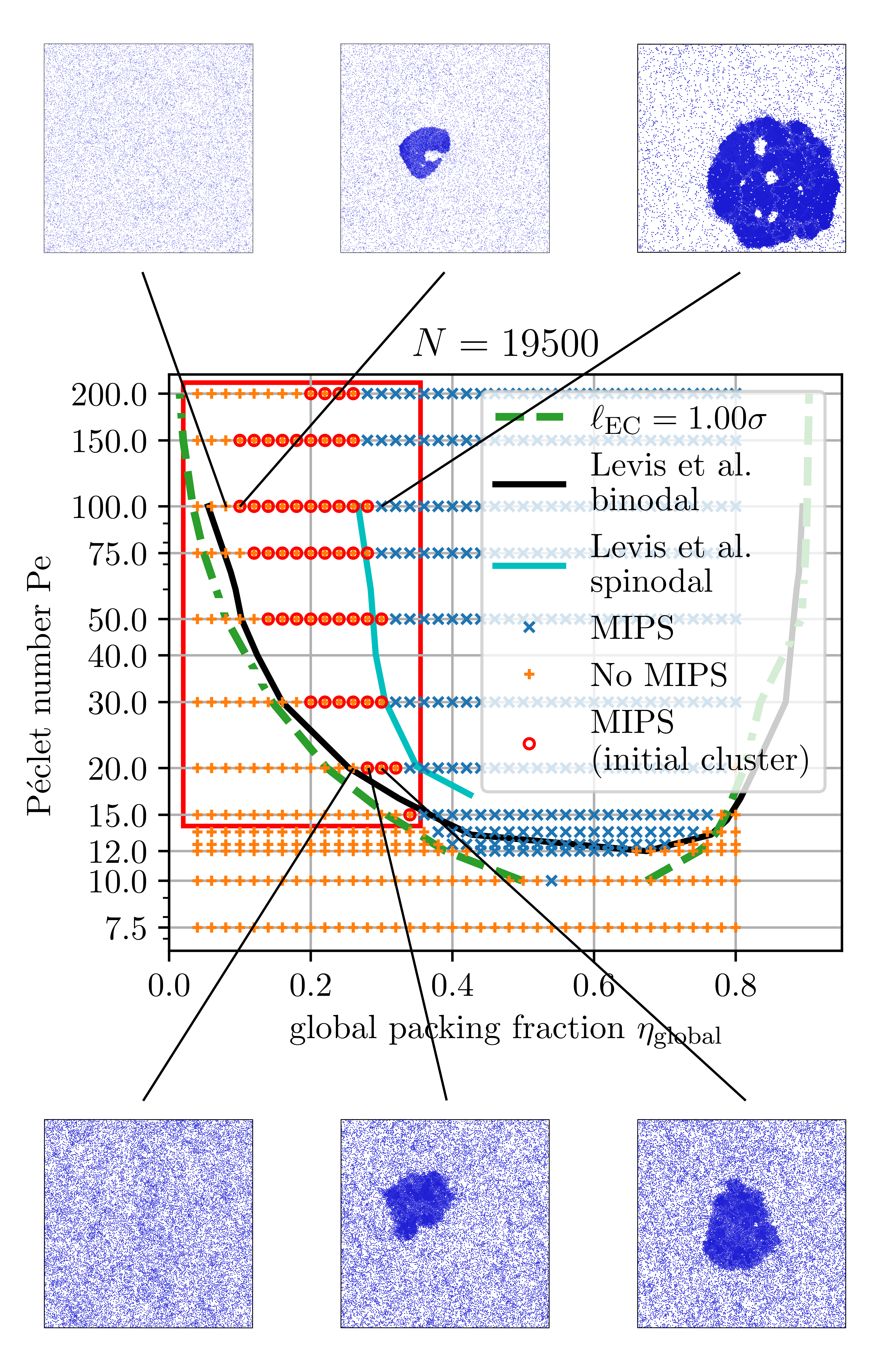}
	\caption{
		Results of additional simulations with phase separated initial state in the phase region 
		enclosed by the red rectangle. The red circles are simulations with final states that 
		show MIPS as indicated for some examples in the insets. This confirms that the boundary 
		between blue and orange points represents a spinodal line.
	}
	\label{fig:mips56}
	\end{center}
\end{figure}

\subsection{Determination of binodal and spinodal lines}

To determine the binodal line, we calculate  the probability density 
distribution $p(\eta_{\mathrm{local}})$ of the
local packing fraction. If MIPS occurs, 
$p(\eta_{\mathrm{local}})$ exhibits two
distinct maxima corresponding to the dilute and dense phase 
(see Fig.\ \ref{fig:histogram}). The binodal lines 
indicate the coexisting densities and are determined by 
averaging the coexisting densities (the positions of the two maxima of $p(\eta_{\mathrm{local}})$) for different 
global packing fractions at the same P\'{e}clet number.

Orange points in Fig.\ \ref{fig:full_mips_phase_diagram} of the main text, 
which lie within the region spanned by the 
binodal line, correspond to global packing fractions, 
where the homogeneous initial state is only metastable. They 
should become unstable to MIPS by strong perturbations, 
such as strong density fluctuations. To check this, some systems
are simulated with $\eta \leq 0.32$, similar to Ref.\,\cite{Levis2017}, 
with an already phase-separated initial state, 
which we prepared by placing a dense and approximately square-shaped 
particle cluster in the center of the system.
In addition, all initial driving forces are aligned in the direction 
of the system center. The results are shown in 
Fig.\ \ref{fig:mips56}, where red circles represent systems 
with phase separated final states (see insets). We find 
that the phase separated state is globally stable, 
while the homogeneous state is only metastable, for most of the 
orange points, which lie between the coexistence lines. 
Small deviations, especially near the binodal, are due to 
finite-size effects cf. to a free energy barrier of 
homogeneous nucleation.

The method to calculate the binodal lines, which was previously explained, 
does not work close to the critical point. This is because in proximity 
of $(\eta_{\mathrm{crit}}, \mathrm{Pe_{crit}})$ one cannot determine the 
peaks related to the dilute and dense phase, as they are barely set apart 
from the rest of the distribution and become virtually indistinguishable 
from small fluctuations of $p(\eta_{\mathrm{local}})$. Of course one can 
deal with these problems by increasing the sample and system size, which
would lead to an increase of the overall computational effort.

\section{Handling of non-periodic boundary conditions and external potentials}

Throughout this work we have focused on periodic boundary conditions, however certain
problems require non-periodicity via soft or hard walls. Since soft walls can be viewed 
as short-ranged external potentials and hard walls are the infinitely steep limit of the 
former, we will briefly address how to incorporate  external one-particle potentials into our
algorithm. 

Before dealing with the kEC algorithm, we should take a closer look at the dynamics of
the new setup. In the presence of an external potential the equation of motion for an
isolated particle takes the form
\begin{equation}
    \dot{\vec{r}}(t) = v_0 \vec{e}(t) + \vec{\xi}(t) - \frac{1}{\Gamma} \vec{F}_\text{ext}(\vec{r}),
    \label{eq:ABP_ext}
\end{equation}
which leads - in the limit of small timesteps - to the mean displacement
\begin{align}
    \langle \Delta \vec{r} \rangle &\approx \Big(v_0 \vec{e} - \frac{1}{\Gamma} \vec{F}_\text{ext}(\vec{r})\Big) \Delta t \nonumber \\
    &= \Big(v_0 \vec{e} - v_\text{ext} \vec{e}_\text{ext} \Big) \Delta t.
    \label{eq:ABP_ext2}
\end{align}

As mentioned in the beginning, a rejection caused by an external potential leads to a 
reflection of the displacement direction at the potentials equipotential surface. There are 
two possibilities of adjusting the mean time per EC 
move for an active particle: One could either calculate the mean move vector 
$\langle \Delta \vec{r}_\text{EC} \rangle$ during one ECMC step for an algorithm that 
handles the active force and external potential separately, or we consider an algorithm that combines both
active and external force into one potential. The first option
keeps up the idea that each interaction can be handled entirely independently in the  ECMC algorithm 
through factorization of the Metropolis filter, but it is
also way more complicated to calculate $\langle \Delta \vec{r}_\text{EC} \rangle$ for such a factorized version of the algorithm. Therefore, we strongly suggest to use the second option. As the previous
analysis proved, the dynamics of a single free ABP is correct in the limit $\ell_\text{EC}/\sigma
\rightarrow 0$. As we expect the same to be true in the presence of an external potential, we
can approximate any potential during a single EC step as a linear potential (this is the same idea 
as the assumption of small timesteps in Eq.\ (\ref{eq:ABP_ext2})). This allows us to calculate
$\langle \vec{r}_\text{EC} \rangle$ analogously to Sec.\ \ref{sec:meantEC} using the combined total force
$F_\text{tot} = \vert F_0 \vec{e} - F_\text{ext} \vec{e}_\text{ext} \vert$ instead of $F_0$ 
and $\vec{e}_\text{tot} = \vec{F}_\text{tot} / F_\text{tot}$ instead of $\vec{e}$. In case of
short-ranged potentials and/or nonlinear potentials this procedure causes the mean time per EC 
step to depend on the position of the particle $\langle \Delta t_\text{EC}(\vec{r}) \rangle$. This
may raise the concern, that the individual particle times may vary greatly. However, because of the shape
of the time function $f(x)$ we expect the particle times to coincide at least in the limit $x \rightarrow 0$ and therefore the limit of vanishing EC lengths. 

\end{document}